# Collaborative Decoding of Interleaved Reed–Solomon Codes and Concatenated Code Designs

Georg Schmidt, *Student Member, IEEE*, Vladimir R. Sidorenko, *Member, IEEE*, and
Martin Bossert *Senior Member, IEEE*
Department of Telecommunications and Applied Information Theory, University of Ulm, Germany
{georg.schmidt,vladimir.sidorenko,martin.bossert}@uni-ulm.de

*Abstract*— Interleaved Reed–Solomon codes are applied in numerous data processing, data transmission, and data storage systems. They are generated by interleaving several codewords of ordinary Reed–Solomon codes. Usually, these codewords are decoded independently by classical algebraic decoding methods. However, by collaborative algebraic decoding approaches, such interleaved schemes allow the correction of error patterns beyond half the minimum distance, provided that the errors in the received signal occur in bursts. In this work, collaborative decoding of interleaved Reed–Solomon codes by multi-sequence shift-register synthesis is considered and analyzed. Based on the framework of interleaved Reed–Solomon codes, concatenated code designs are investigated, which are obtained by interleaving several Reed–Solomon codes, and concatenating them with an inner block code.

## I. INTRODUCTION

In recent years, code designs with *interleaved Reed–Solomon* (IRS) codes were the topic of several scientific publications. They are investigated by different authors like Krachkovsky, Lee, and Garg [1], [2], [3], Bleichenbacher, Kiayias, and Yung [4], Brown, Minder and Shokrollahi [5], [6], Justesen, Thommesen, and Høholdt [7], as well as Parvaresh and Vardy [8]. Moreover, IRS codes are considered in [9], [10], [11], and other publications. Interleaved Reed–Solomon codes are mainly considered in applications where error bursts occur, since IRS codes are most effective if correlated errors affect all words of the interleaved scheme simultaneously. In [3], [7], and [10], it is suggested to consider IRS codes also as outer codes in concatenated code designs, since even for channel models inducing statistically independent random errors, the decoder for the inner code will usually create correlated burst errors at the input of the decoder for the outer codes. For decoding IRS codes, Bleichenbacher, Kiayias, and Yung [4] propose an algorithm based on the *Welch–Berlekamp* approach. Parvaresh, and Vardy [8] consider IRS decoding in the context of multivariate polynomial interpolation, which yields a quite powerful list decoding algorithm.

In this paper, we consider IRS codes in a rather general way. More precisely we consider IRS codes consisting of $l$ Reed–Solomon codes $\mathcal{A}^{(1)}, \mathcal{A}^{(2)}, \ldots, \mathcal{A}^{(l)}$ of length $N$ and dimensions $K^{(1)}, K^{(2)}, \ldots, K^{(l)}$. We take a codeword from each code $\mathcal{A}^{(\ell)}$, $\ell = 1, \ldots, l$, and arrange them row-wise into a matrix like depicted in Fig. 1. We call the set of matrices obtainable in this way an IRS code. If $\mathcal{A}^{(1)} = \mathcal{A}^{(2)} = \cdots = \mathcal{A}^{(l)}$, we call the code *homogeneous* IRS code, if the codes are different, we call the resulting IRS code *heterogeneous*.

Most previous publications consider only homogeneous IRS codes. Heterogeneous constructions have first been considered in [3] (without calling them heterogeneous IRS codes), and some of their properties have been investigated in [12]. Heterogeneous IRS codes may be interesting, e.g. when used as outer codes of *generalized concatenated codes* introduced and described by Blokh and Zyablov [13] and by Zinoviev [14]. Another interesting application of heterogeneous IRS codes is the decoding of a single Reed–Solomon code beyond half the minimum distance like described in [15], where the problem of decoding a single low-rate Reed–Solomon code is transformed into the problem of decoding a heterogeneous IRS code.

We propose a method for collaborative decoding of both homogeneous and heterogeneous IRS codes, which is based on multi-sequence shift-register synthesis. To analyze the behavior of this decoder, we derive bounds on the decoding error and decoding failure probability. These bounds allow for estimating of the gain, which can be achieved by collaborative decoding in comparison to decoding the $l$ Reed–Solomon codes independently up to half the minimum distance by a standard *Bounded Minimum Distance* (BMD) decoder.

If one column of the arrangement shown by Fig. 1 is corrupted by a burst error, i.e., an error which affects a complete column of the arrangement, the $l$ Reed–Solomon codewords may have an erroneous symbol at the same position. Hence, a collaborative decoding strategy can be applied, which locates the errors jointly in all Reed–Solomon codewords instead of locating them independently in the several words. This allows for uniquely locating up to $t$ errors, in many cases even if $t$ is larger than half the minimum distance of the Reed–Solomon

The work of Vladimir R. Sidorenko and Georg Schmidt is supported by Deutsche Forschungsgemeinschaft (DFG), Germany, under project BO 867/14, and BO 867/15 respectively. V. R. Sidorenko is on leave from IITP, RAS RU. Some of the results presented in this paper have also been presented on the 2005 IEEE ITSOC Information Theory Workshop in Rotorua, New Zealand, and the 10th International Workshop on Algebraic and Combinatorial Coding Theory (ACCT-10) 2006 in Zvenigorod, Russia. This work has been submitted to the IEEE for possible publication. Copyright may be transferred without notice, after which this version will be superseded.

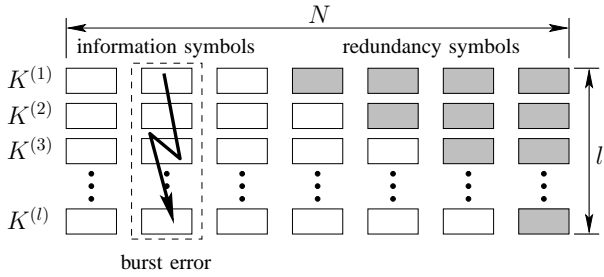

Fig. 1. Interleaved Reed–Solomon code.

code with the largest dimension.

Algebraic decoding of a single Reed–Solomon codeword can efficiently be performed by the *Berlekamp–Massey algorithm* [16], [17], which is based on a single-sequence shift-register synthesis approach. It is mentioned in [2] and [3], that decoding of interleaved Reed–Solomon codes can be performed on the basis of a multi-sequence shift-register synthesis algorithm. Such an algorithm is described by Feng and Tzeng in [18] and [19]. For homogeneous IRS codes, this algorithm provides an effective method for collaborative decoding. However, it is shown in [20] and [21] that the *Feng–Tzeng algorithm* does not always work correctly for sequences of varying length. Thus, it cannot be applied for the heterogeneous case, since the different dimensions of the codes $\mathcal{A}^{(1)}, \mathcal{A}^{(2)}, \ldots, \mathcal{A}^{(l)}$ result in syndrome sequences of different lengths. Hence, we use the multi-sequence shift-register synthesis algorithm proposed in [20] to decode both, homogeneous and heterogeneous IRS codes. The complexity of this shift-register based approach is similar to the complexity of applying the Berlekamp–Massey algorithm to all words of the interleaved Reed–Solomon code independently.

In the following sections, we describe the basic principles behind collaborative IRS decoding. Based on this, we derive the maximum correction radius. Furthermore, we discuss how the code properties and the decoding performance is influenced by the IRS code design, i.e., by the choice of the parameters of $\mathcal{A}^{(1)}, \mathcal{A}^{(2)}, \ldots, \mathcal{A}^{(l)}$. Moreover, we derive bounds on the failure and error probability, which allow us to estimate the performance gain, which is achievable by collaborative decoding. On the basis of these results, we investigate concatenated code designs with outer IRS codes and inner block codes, and derive bounds on the overall decoding performance. These considerations are supplemented by *Monte Carlo simulations*, to assess the quality of our bounds.

## II. INTERLEAVED REED–SOLOMON CODES

As already mentioned, the codewords of an IRS code are matrices whose rows are the codewords of the Reed–Solomon codes $\mathcal{A}^{(1)}, \mathcal{A}^{(2)}, \ldots, \mathcal{A}^{(l)}$. Before we formally define IRS codes, we briefly introduce classical Reed–Solomon codes, basically to establish the notation we use in the following.

Reed–Solomon codes can conveniently be defined in the frequency domain, i.e., by the Fourier Transform of their codewords. For this purpose, we consider the following definition of the Discrete Fourier Transform over finite fields:

**Definition 1 (Discrete Fourier Transform (DFT) over $\mathbb{F}_q$)**
Let $p(x) = p_0 + p_1 x + \cdots + p_{n-1} x^{n-1}$ be a polynomial of degree $\deg(p(x)) \leq n-1$ with coefficients from $\mathbb{F}_q$. Further, let $\alpha \in \mathbb{F}_q$ be some element of order $n$. Then, the polynomial

$$P(x) = \mathscr{F}(p(x)) = P_0 + P_1 x + \cdots + P_{n-1} x^{n-1}$$

whose coefficients are calculated by $P_i = p(\alpha^i)$ is called the Discrete Fourier Transform of $p(x)$ over the field $\mathbb{F}_q$.

The inverse of the Discrete Fourier Transform pursuant to Definition 1 can be calculated in the well-known way as follows: Let $P(x)$ be a polynomial of $\deg(P(x)) \leq n-1$, and denote by

$$p(x) = \mathscr{F}^{-1}(P(x)) = p_0 + p_1 x + \cdots + p_{n-1} x^{n-1}$$

the *Inverse Discrete Fourier Transform*. Then, the coefficients $p_i$ of $p(x)$ are calculated by $p_i = n^{-1} P(\alpha^{-i})$.

In the terminology of the Fourier Transform, we formally call $p(x)$ *time domain polynomial* and $P(x)$ the *frequency domain polynomial* or the *spectrum* of $p(x)$. We now describe Reed–Solomon codes based on the Discrete Fourier Transform. The following definition does not describe the class of Reed–Solomon codes in its most general way. However, with regard to a concise notation, it is adequate for our purposes:

**Definition 2 (Reed–Solomon (RS) code)**
Let

$$\{A(x)\} = \left\{ \sum_{i=0}^{K-1} A_i x^i, \ A_i \in \mathbb{F}_q \right\}$$

be the set of all polynomials of $\deg(A(x)) < K$ with coefficients $A_i$ from $\mathbb{F}_q$, and let $\alpha \in \mathbb{F}_q$ be some element of order $N < q$. Then, a Reed–Solomon code $\mathcal{A} = \mathcal{RS}(q; N, K, D)$ of length $N$, dimension $K$, and minimum Hamming distance $D = N - K + 1$ can be defined as the set of polynomials

$$\mathcal{A} \triangleq \left\{ a(x) = \mathscr{F}^{-1}(A(x)) \mid A(x) \in \{A(x)\} \right\}.$$

The codewords of $\mathcal{A}$ are represented by the polynomials $a(x) = a_0 + a_1 x + \cdots + a_{n-1} x^{n-1}$, or alternatively by the n-tuples $\boldsymbol{a} = (a_0, a_1, \ldots, a_{n-1})$.

Reed–Solomon codes fulfill the *Singleton Bound* with equality, and are therefore *Maximum Distance Separable* (MDS). This means that from any set of $K$ correct symbols, a Reed–Solomon codeword can uniquely be reconstructed.

An interleaved Reed–Solomon code is now obtained by taking $l$ Reed–Solomon codes according to Definition 2 and grouping them row-wise into a matrix:

**Definition 3 (Interleaved Reed–Solomon (IRS) code)**
Let $\mathcal{A}^{(\ell)} = \mathcal{RS}(q; N, K^{(\ell)}, D^{(\ell)})$, $\ell = 1, \ldots, l$, be $l$ Reed–Solomon codes of length $N$ according to Definition 2. Then, an interleaved Reed–Solomon code is the set of matrices

$$\boldsymbol{\mathcal{A}} \triangleq \left\{ \begin{pmatrix} \boldsymbol{a}^{(1)} \\ \boldsymbol{a}^{(2)} \\ \vdots \\ \boldsymbol{a}^{(l)} \end{pmatrix}, \ \boldsymbol{a}^{(\ell)} \in \mathcal{A}^{(\ell)}, \ \ell = 1, \ldots, l \right\}.$$

*If all $l$ Reed–Solomon codes are equivalent, i.e., $\mathcal{A}^{(1)} = \mathcal{A}^{(2)} = \cdots = \mathcal{A}^{(l)} = \mathcal{A}$, the IRS code is called* homogeneous. *Otherwise, we say that the IRS code is* heterogeneous.

Instead of considering an IRS codeword as $l \times N$ matrix $\mathbf{A}$ with elements from the extension field $\mathbb{F}_q$, we can also consider it as a row vector $\mathbf{a}$ with elements from the extension field $\mathbb{F}_{q^l}$. In this extension field representation, an IRS code is a code of length $N$, cardinality $\prod_{\ell=1}^{l} q^{k^{(\ell)}}$, and minimum distance $\min_{1 \leq \ell \leq l} \left\{ N - K^{(\ell)} \right\} + 1$ over $\mathbb{F}_{q^l}$. Hence, for the homogeneous case, i.e., if $\mathcal{A}^{(\ell)} = \mathcal{RS}(q; N, K, D)$, $\ell = 1, \ldots, l$, we obtain a code of length $N$, dimension $K$, and minimum distance $D = N - K + 1$ over $\mathbb{F}_{q^l}$. This means that homogeneous IRS codes fulfill the Singleton Bound with equality and are therefore MDS. However, since the minimum distance of a heterogeneous IRS code is determined by the weakest Reed–Solomon code, i.e., the Reed–Solomon code with the smallest minimum distance, heterogeneous IRS codes are generally not MDS.

## III. COLLABORATIVE DECODING OF INTERLEAVED REED–SOLOMON CODES

Before we explain the concept of collaborative interleaved Reed–Solomon decoding, we consider the conventional case of a single Reed–Solomon code $\mathcal{A} = \mathcal{RS}(q; N, K, D)$. Decoding is usually performed by an algebraic *Bounded Minimum Distance* (BMD) decoder, which corrects all errors with weights up to half the minimum code distance.

### A. BMD Decoding up to Half the Minimum Distance

Assume that a codeword $a(x) \in \mathcal{A}$ is transmitted over a noisy channel, which adds an error polynomial $e(x) = e_0 + e_1 x + \cdots + e_{N-1} x^{N-1}$ over $\mathbb{F}_q$, so that we observe the word $r(x) = a(x) + e(x)$ at the output of the channel. If $e(x) \equiv 0$, the Discrete Fourier Transform $R(x) = \mathscr{F}(r(x)) = R_0 + R_1 x + \cdots + R_{N-1} x^{N-1}$ will be a polynomial of degree smaller than $K$ due to Definition 2, i.e., the coefficients $R_K, \ldots, R_{N-1}$ will be zero. However, if $e(x) \notin \mathcal{A}$, we obtain $\deg(R(x)) \geq K$, which means that some of the coefficients $R_K, \ldots, R_{N-1}$ are non-zero. These coefficients only depend on the transformed error $E(x) = \mathscr{F}(e(x)) = E_0 + E_1 x + \cdots + E_{N-1} x^{N-1}$. Hence, we can consider the coefficients $R_K, \ldots, R_{N-1}$ as syndrome coefficients and denote them by $S_j = R_{K+j} = E_{K+j}$, $j = 0, \ldots, N - K - 1$.

To correct $t$ errors, the standard approach for algebraic Reed–Solomon decoding is to define the polynomial $\lambda(x) = \lambda_0 + \lambda_1 x + \cdots + \lambda_{N-1} x^{N-1}$, such that it has a zero coefficient $\lambda_j = 0$, whenever the corresponding coefficient $e_j$ of the error polynomial $e(x)$ is non-zero. For all error free positions, i.e., for all positions for which $e_j = 0$, the corresponding coefficient $\lambda_j$ is defined to be non-zero. Consequently, $\lambda_j \cdot e_j = 0$ holds for all positions $j = 0, \ldots, N-1$. Due to the properties of the Discrete Fourier Transform, this relation is transformed into

$$\Lambda(x) \cdot E(x) \equiv 0 \mod x^N - 1. \quad (1)$$

The spectrum $\Lambda(x)$ is a polynomial of degree $t$, whose roots $\alpha^{-j_1}, \alpha^{-j_2}, \ldots, \alpha^{-j_t}$ correspond to the locations $j_1, j_2, \ldots, j_t$ of the erroneous symbols. Therefore, $\Lambda(x) = \Lambda_0 + \Lambda_1 x + \cdots + \Lambda_t x^t$ is called *error locator polynomial*. An error locator polynomial was first applied by Peterson [22] for decoding BCH codes.

Since the roots of the error locator polynomial are not modified by multiplying $\Lambda(x)$ by a constant factor, and since $\Lambda_0 \neq 0$, $\Lambda(x)$ can always be normalized w.l.o.g. in such a way that $\Lambda_0 = 1$. Equation (1) forms a linear system of $N$ equations. In this system, $t$ equations only depend on the $M = N - K$ known syndrome coefficients $S_0, \ldots, S_{M-1}$ and the unknown coefficients $\Lambda_1, \ldots, \Lambda_t$. With these $t$ equations, we write the matrix equation

$$\underbrace{\begin{pmatrix} S_0 & S_1 & \ldots & S_{t-1} \\ S_1 & S_2 & \ldots & S_t \\ \vdots & \vdots & & \vdots \\ S_{M-t-1} & S_{M-t} & \ldots & S_{M-2} \end{pmatrix}}_{\boldsymbol{S}} \underbrace{\begin{pmatrix} \Lambda_t \\ \Lambda_{t-1} \\ \vdots \\ \Lambda_1 \end{pmatrix}}_{\boldsymbol{\Lambda}} = \underbrace{\begin{pmatrix} -S_t \\ -S_{t+1} \\ \vdots \\ -S_{M-1} \end{pmatrix}}_{\boldsymbol{T}}, \quad (2)$$

which is a linear system of $N - K - t$ equations and $t$ unknowns. Hence, (2) cannot have a unique solution, if $t > \frac{N-K}{2}$ and we are never able to correct more than $\lfloor \frac{N-K}{2} \rfloor$ errors. If $t \leq \lfloor \frac{N-K}{2} \rfloor$, solving (2) yields a unique error locator polynomial, and hence also the locations of the erroneous symbols. Basically, (2) can be solved by standard methods from linear algebra. This approach, also known as *Peterson Algorithm*, is described in [23]. However, today the Peterson Algorithm is of minor practical relevance since there exist algorithms which utilize the structure of (2) to solve the system of equations in a more efficient way.

To understand how the structure of (2) can be exploited, we observe that we are able to state (2) in the form of the linear recursion

$$S_i = -\sum_{j=1}^{t} \Lambda_j S_{i-j}, \quad i = t, \ldots, N - K^{(\ell)} - 1 \quad (3)$$

of length $t$. In this way, the problem of calculating $\Lambda(x)$ is transformed to the problem of finding the smallest integer $t$ and the connection weights $\Lambda_1, \ldots, \Lambda_t$ for recursively generating the syndrome sequence $\mathcal{S} = \{S_i\}_{i=0}^{N-K-1}$. This problem is equivalent to the problem of determining the shortest possible linear feedback shift-register, which is capable of generating the syndrome sequence $\mathcal{S}$. This single-sequence shift-register synthesis problem is solved very efficiently by the Berlekamp–Massey algorithm [16], [17].

When $\Lambda(x)$ is determined, and hence the erroneous positions are known, the most difficult part of decoding is accomplished. If the error locations are known, the error values can always be uniquely determined, as long as $t \leq N - K$. Error evaluation can be performed using several standard techniques like *Recursive Extension* [24] or the *Forney algorithm* [25].

### B. Collaborative Error Location

Now, we consider an interleaved Reed–Solomon code pursuant to Definition 3. We assume that we observe a received

word

$$R = \begin{pmatrix} r^{(1)} \\ r^{(2)} \\ \vdots \\ r^{(l)} \end{pmatrix} = \underbrace{\begin{pmatrix} a^{(1)} \\ a^{(2)} \\ \vdots \\ a^{(l)} \end{pmatrix}}_{A} + \underbrace{\begin{pmatrix} e^{(1)} \\ e^{(2)} \\ \vdots \\ e^{(l)} \end{pmatrix}}_{E},$$

where the $\ell$th row of $R$ consists of the coefficients of the received word $r^{(\ell)}(x) = a^{(\ell)}(x) + e^{(\ell)}(x)$. Each error polynomial $e^{(\ell)}(x)$ has $t^{(\ell)} \leq t$ non-zero coefficients located at the indices $\{i_1, i_2, \ldots, i_{t^{(\ell)}}\} \subseteq \{j_1, j_2, \ldots, j_t\}$. Hence, $R = A + E$ can be written as sum of a codeword $A \in \mathcal{A}$ and an error matrix $E$ with exactly $t$ non-zero columns. For each row in $R$, we are able to calculate a syndrome sequence $\mathcal{S}^{(\ell)} = \{S_i^{(\ell)}\}_{i=0}^{N-K^{(\ell)}-1}$. However, since we allow the codes $\mathcal{A}^{(\ell)}$ to have different dimensions $K^{(\ell)}$, the calculated syndrome sequences may be of different lengths. Nevertheless, as long as $t < N - K^{(\ell)}$, we are able to create the $(N - K^{(\ell)} - t) \times t$ matrix

$$S^{(\ell)} = \begin{pmatrix} S_0^{(\ell)} & S_1^{(\ell)} & \cdots & S_{t-1}^{(\ell)} \\ S_1^{(\ell)} & S_2^{(\ell)} & \cdots & S_t^{(\ell)} \\ \vdots & \vdots & & \vdots \\ S_{N-K^{(\ell)}-t-1}^{(\ell)} & S_{N-K^{(\ell)}-t}^{(\ell)} & \cdots & S_{N-K^{(\ell)}-2}^{(\ell)} \end{pmatrix},$$

and a vector

$$T^{(\ell)} = \begin{pmatrix} -S_t^{(\ell)} \\ -S_{t+1}^{(\ell)} \\ \vdots \\ -S_{N-K^{(\ell)}-1}^{(\ell)} \end{pmatrix}$$

for each sequence $\mathcal{S}^{(\ell)}$, $\ell = 1, \ldots, l$. We use the matrices $S^{(1)}, S^{(2)}, \ldots, S^{(l)}$, and the vectors $T^{(1)}, T^{(2)}, \ldots, T^{(l)}$ to state the linear system of equations

$$\underbrace{\begin{pmatrix} S^{(1)} \\ S^{(2)} \\ \vdots \\ S^{(l)} \end{pmatrix}}_{S_l} \cdot \underbrace{\begin{pmatrix} \Lambda_t \\ \Lambda_{t-1} \\ \vdots \\ \Lambda_1 \end{pmatrix}}_{\Lambda} = \underbrace{\begin{pmatrix} T^{(1)} \\ T^{(2)} \\ \vdots \\ T^{(l)} \end{pmatrix}}_{T_l} \quad (4)$$

with $t$ unknowns. Note that if $t = N - K^{(\ell)}$ for some Reed–Solomon code $\mathcal{A}^{(\ell)}$, we are still able to state a system of equations similar to (4), simply by skipping the corresponding matrix $S^{(\ell)}$ and the corresponding vector $T^{(\ell)}$. Moreover, it is senseless to consider the case $t > N - K^{(\ell)}$, since we are not able to uniquely reconstruct the Reed–Solomon codeword $a^{(\ell)}(x)$ from less than $K^{(\ell)}$ uncorrupted symbols.

### C. Error Location by Multi-Sequence Shift-Register Synthesis

The basic structure of (4) is similar to the structure of (2). Thus, similar as before, we create the set of linear recursions

$$S_i^{(\ell)} = -\sum_{j=1}^{t} \Lambda_j S_{i-j}^{(\ell)}, i = t, \ldots, N - K^{(\ell)} - 1, \quad \ell = 1, \ldots, l, \quad (5)$$

of length $t$. All $l$ linear recursions use the same connection weights $\Lambda_1, \ldots, \Lambda_t$ to combine the syndrome coefficients of the $l$ received words. Hence, the error locator polynomial $\Lambda(x)$ can be calculated by finding the smallest integer $t$ and the connection weights $\Lambda_1, \ldots, \Lambda_t$ for recursively generating all $l$ different syndrome sequences $\mathcal{S}^{(\ell)} = \{S_i\}_{i=0}^{N-K^{(\ell)}-1}$. This is equivalent to synthesizing the shortest linear feedback shift-register capable of generating the $l$ syndrome sequences $\mathcal{S}^{(1)}, \ldots, \mathcal{S}^{(l)}$. For a homogeneous Reed–Solomon code, i.e., if all $l$ sequences have the same length, this multi-sequence shift-register problem can be solved by the Feng-Tzeng algorithm [18], [19]. However, for a heterogeneous IRS code, the sequences $\mathcal{S}^{(1)}, \ldots, \mathcal{S}^{(l)}$ may be of different length. It is demonstrated in [20] and [21] that the Feng–Tzeng algorithm does not always yield the shortest shift-register for varying length sequences. Hence, we apply the varying length shift-register synthesis algorithm proposed in [20] to calculate $\Lambda(x)$ for both homogeneous and heterogeneous IRS codes. A pseudo-code description of this algorithm is given by Algorithm 1.

---

**Algorithm 1:** Shift-Register Synthesis Algorithm from [20]

**input**: $\mathcal{S}^{(\ell)} = \{S_i^{(\ell)}\}_{i=0}^{N-K^{(\ell)}-1}$, $\ell = 1, \ldots, l$

$M \leftarrow \max_{1 \leq \ell \leq l} \{N - K^{(\ell)}\}$
$M^{(\ell)} \leftarrow N - K^{(\ell)}$, $\ell = 1, \ldots, l$
$t \leftarrow 0$, $\Lambda(x) \leftarrow 1$
$m^{(\ell)} \leftarrow M - M^{(\ell)}$, $t^{(\ell)} \leftarrow 0$, for $\ell = 1, \ldots, l$
$\Lambda^{(\ell)}(x) \leftarrow 0$, $\Delta^{(\ell)} \leftarrow 1$, for $\ell = 1, \ldots, l$
**for** *each $m$ from 0 to $M - 1$* **do**
  **for** *each $\ell$ from 1 to $l$* **do**
    **if** $m - t > M - M^{(\ell)}$ **then**
      $\Delta \leftarrow S_{m-M+M^{(\ell)}}^{(\ell)} + \sum_{i=1}^{t} \Lambda_i S_{m-M+M^{(\ell)}-i}^{(\ell)}$
      **if** $\Delta \neq 0$ **then**
        **if** $m - m^{(\ell)} \leq t - t^{(\ell)}$ **then**
          $\Lambda(x) \leftarrow \Lambda(x) - \frac{\Delta}{\Delta^{(\ell)}} \Lambda^{(\ell)}(x) x^{m-m^{(\ell)}}$
        **else**
          $\tilde{t} \leftarrow t$, $\tilde{\Lambda}(x) \leftarrow \Lambda(x)$
          $\Lambda(x) \leftarrow \Lambda(x) - \frac{\Delta}{\Delta^{(\ell)}} \Lambda^{(\ell)}(x) x^{m-m^{(\ell)}}$
          $t \leftarrow m - (m^{(\ell)} - t^{(\ell)})$
          $t^{(\ell)} \leftarrow \tilde{t}$, $\Lambda^{(\ell)}(x) \leftarrow \tilde{\Lambda}(x)$
          $\Delta^{(\ell)} \leftarrow \Delta$, $m^{(\ell)} \leftarrow m$

**output**: $t$, $\Lambda(x)$

---

Applying Algorithm 1 to the $l$ syndromes $\mathcal{S}^{(1)}, \ldots, \mathcal{S}^{(l)}$ yields a polynomial $\Lambda(x)$ and a shift register length $t$. However, by the definition of the error locator polynomial, $\Lambda(x)$ is only a valid error locator polynomial, if it has exactly $t$ distinct roots. Hence, we accept a $\Lambda$-polynomial obtained from Algorithm 1 only, if it conforms to the following definition:

**Definition 4 ($t$-valid $\Lambda$-polynomial)**
*A polynomial $\Lambda(x)$ over $\mathbb{F}_q$ is called $t$-valid, if it is a polynomial of degree $t$ and possesses exactly $t$ distinct roots in $\mathbb{F}_q$.*

Once a $t$-valid error locator polynomial $\Lambda(x)$ is successfully calculated for an IRS code, error evaluation can be performed independently for all $l$ Reed–Solomon codewords by the standard techniques also used for classical BMD decoding.

Using Algorithm 1 for calculating $\Lambda(x)$ yields the same computational complexity as locating the errors in the $l$ codewords independently by the Berlekamp–Massey algorithm. Consequently, we are able to increase the error correction radius above half the minimum distance of the Reed–Solomon code with the smallest distance, without increasing the decoding complexity.

Altogether, our collaborative decoding strategy for IRS codes consists of four steps. First, we calculate $l$ syndromes $\mathcal{S}^{(1)}, \ldots, \mathcal{S}^{(l)}$, next, we synthesize an error locator polynomial $\Lambda(x)$, then we check whether this polynomial is $t$-valid, and whether $t$ is smaller than some maximum error correcting radius $t_{\max}$, which will be specified later. If we have a $t$-valid $\Lambda$-polynomial, we calculate values of the errors $e^{(\ell)}(x)$ independently for all received words $r^{(\ell)}(x)$, and obtain estimates

$$\hat{a}^{(\ell)}(x) = r^{(\ell)}(x) - e^{(\ell)}(x)$$

for $\ell = 1, \ldots, l$. If $\Lambda(x)$ is not $t$-valid, we get a decoding failure. The complete decoding algorithm is summarized by Algorithm 2.

---

**Algorithm 2:** Collaborative IRS Decoder

**input**: received word $\boldsymbol{R} = \begin{pmatrix} \boldsymbol{r}^{(1)} \\ \vdots \\ \boldsymbol{r}^{(l)} \end{pmatrix}$

use DFT to calculate syndromes $\mathcal{S}^{(1)}, \ldots, \mathcal{S}^{(l)}$
synthesize $t$, $\Lambda(x)$ by Algorithm 1
**if** $t \leq t_{\max}$ **and** $\Lambda(x)$ *is $t$-valid* **then**
    **for** *each $\ell$ from 1 to $l$* **do**
        evaluate errors, and calculate $\boldsymbol{e}^{(\ell)}$
        calculate $\hat{\boldsymbol{a}}^{(\ell)} = \boldsymbol{r}^{(\ell)} - \boldsymbol{e}^{(\ell)}$
**else**
    decoding failure

**output**: $\widehat{\boldsymbol{A}} = \begin{pmatrix} \hat{\boldsymbol{a}}^{(1)} \\ \vdots \\ \hat{\boldsymbol{a}}^{(l)} \end{pmatrix} \in \mathcal{A}$ or decoding failure

---

Generally, depending on the errors added by the channel, Algorithm 2 may yield three different results:

1) The algorithm may obtain a correct result, i.e., $a^{(\ell)}(x) \equiv \hat{a}^{(\ell)}(x) \, \forall \ell = 1, \ldots, l$.
2) The algorithm may obtain an erroneous result, i.e., $\exists \ell \in [1, \ldots, l] : a^{(\ell)}(x) \not\equiv \hat{a}^{(\ell)}(x)$.
3) The algorithm may not yield a result at all, i.e., it may yield a decoding failure.

All three events occur with a certain probability. In the following, we denote the probability for a correct decision by $P_c$, the probability for an erroneous decision by $P_e$, and the probability for a decoding failure by $P_f$. Hence, the probability $P_w$ for obtaining a wrong decoding result is calculated by $P_w = 1 - P_c = P_e + P_f$.

## D. Joint Error and Erasure Correction

In [11], an algorithm for joint error and erasure decoding of IRS codes is proposed. Like the algorithm presented in [4] for decoding errors only, the algorithm from [11] uses a *Welch-Berlekamp* approach, which is based on solving a linear system of equations. We briefly explain how Algorithm 2 can be modified to allow joint error and erasure correction based on shift-register synthesis.

Technically, erasures occur if the receiver is not able to detect any meaningful symbol at the output of the channel. This means that the decoder does not know the value, but the position of an erased symbol. Formally, we define an erasure as follows: Let $\xi$ denote an erased symbol, and define it to be a special symbol $\xi \notin \mathbb{F}_q$. Moreover, let the addition of $\alpha \in \mathbb{F}_q$ and $\xi$ be defined by $\alpha + \xi = \xi + \alpha = \xi$. Now, assume that we observe a received word

$$\boldsymbol{R} = \boldsymbol{A} + \boldsymbol{E} + \boldsymbol{\mathcal{E}},$$

where the rows of the matrix

$$\boldsymbol{\mathcal{E}} = \begin{pmatrix} \boldsymbol{\epsilon}^{(1)} \\ \boldsymbol{\epsilon}^{(2)} \\ \vdots \\ \boldsymbol{\epsilon}^{(l)} \end{pmatrix}$$

are the coefficients of the erasure polynomials $\epsilon^{(\ell)}(x) = \epsilon_0^{(\ell)} + \epsilon_1^{(\ell)} x + \cdots + \epsilon_{n-1}^{(\ell)} x^{n-1}$ over $\{0, \xi\}$, $\ell = 1, \ldots, l$. Each nonzero element in $\boldsymbol{\mathcal{E}}$ indicates an erased symbol. In contrast to errors, the erased symbols are detectable in the received word $\boldsymbol{R}$, which means that they do not have to be located, only their values have to be determined. For this reason, we do not require the erasures to occur in column bursts.

To perform joint error and erasure decoding for IRS codes, we proceed like described in [26] and define an *erasure locator polynomial*

$$\Psi^{(\ell)}(x) \triangleq \left(1 - \alpha^{i_1^{(\ell)}} x\right)\left(1 - \alpha^{i_2^{(\ell)}} x\right) \cdots \left(1 - \alpha^{i_{\vartheta_\ell}^{(\ell)}} x\right) =$$
$$= 1 + \Psi_1^{(\ell)} x + \Psi_2^{(\ell)} x^2 + \cdots + \Psi_{\vartheta_\ell}^{(\ell)} x^{\vartheta_\ell}$$

for each of the rows $\ell = 1, \ldots, l$ in our IRS code. The roots $\alpha^{-i_1^{(\ell)}}, \alpha^{-i_2^{(\ell)}}, \ldots, \alpha^{-i_{\vartheta_\ell}^{(\ell)}}$ indicate the positions of the erasures in the erasure polynomial $\epsilon^{(\ell)}(x)$, i.e., the coefficients of $\epsilon^{(\ell)}(x)$ at the positions $i_1^{(\ell)}, i_2^{(\ell)}, \ldots, i_{\vartheta_\ell}^{(\ell)}$ are equal to $\xi$, and all other coefficients are zero. Now we consider the received polynomial $r^{(\ell)}(x)$ whose coefficients are the $\ell$th row of $\boldsymbol{R}$. From $r^{(\ell)}(x)$ we create the polynomial $\bar{r}^{(\ell)}(x)$ by replacing all erasure elements by an arbitrary field element. Then, we calculate $\bar{R}^{(\ell)}(x) = \mathscr{F}\left(\bar{r}^{(\ell)}(x)\right)$ and $\breve{R}^{(\ell)}(x) = \bar{R}^{(\ell)}(x)\Psi(x)$. It is explained in [26] that $\breve{R}^{(\ell)}(x) = \breve{R}_0^{(\ell)} + \breve{R}_1^{(\ell)} x + \cdots + \breve{R}_{n-1}^{(\ell)} x^{n-1}$ is the spectrum of a modified received word $\breve{r}^{(\ell)}(x)$ with errors at the same positions as $r^{(\ell)}(x)$. Moreover, the last $N - K^{(\ell)} - \vartheta_\ell$ coefficients of $\breve{R}^{(\ell)}(x)$ are only influenced by the errors, so that we are able to obtain the modified syndromes $\breve{\mathcal{S}}^{(\ell)} = \{\breve{S}_i^{(\ell)}\}_{i=0}^{N-K^{(\ell)}-1-\vartheta_\ell}$, where $\breve{S}_i^{(\ell)} = \breve{R}_{K^{(\ell)}+\vartheta_\ell+i}^{(\ell)}$, $i = 0, \ldots, N - K^{(\ell)} - \vartheta_\ell - 1$, $\ell = 1, \ldots, l$. This means that each erasure symbol in the row $\ell$ reduces the length of the usable syndrome $\breve{\mathcal{S}}^{(\ell)}$ by one. In this way, we obtain $l$ (varying length) syndrome sequences, which we feed into Algorithm 1 to calculate a pair $t$, $\Lambda(x)$. Once, a $t$-valid polynomial $\Lambda(x)$ is calculated and $t \leq t_{\max}$, we proceed like described in [26] to obtain the error and erasure values independently for each row, using the *error and erasure locator polynomial* $\Omega^{(\ell)}(x) = \Psi^{(\ell)}(x)\Lambda(x)$.

In some sense, error and erasure decoding of an IRS code is equivalent to error only decoding of a heterogeneous IRS code, since in both cases we process varying length syndrome sequences. For this reason, we do not further consider the error and erasure case explicitly, but keep in mind that it is closely related to the case of heterogeneous IRS codes.

## IV. Error Correcting Radius

If the number of errors $t$ is smaller than half the minimum distance of the Reed–Solomon code with the smallest redundancy, we are always able to correct all errors by solving (3) separately for all $l$ Reed–Solomon words using standard techniques like the Berlekamp–Massey algorithm. Certainly, it is important that Algorithm 2 is also capable of decoding all errors in this case. To prove that Algorithm 2 actually has this property, we first consider the following lemma:

**Lemma 1** *Let $v$, and $w$ be two vectors in $\mathbb{F}_q^n$, such that the Hamming weights of $v$ and $w$ satisfy $\mathrm{wt}(v) < q-1$ or $\mathrm{wt}(w) < q-1$. Then, there always exists an element $\beta \in \mathbb{F}_q$, $\beta \neq 0$, such that the support of $v + \beta w$ satisfies*

$$\mathrm{supp}(v + \beta w) = \mathrm{supp}(v) \cup \mathrm{supp}(w) . \tag{6}$$

*Proof:* Since $\mathrm{wt}(v) < q-1$, or $\mathrm{wt}(w) < q-1$ the cardinality of the intersection $\mathcal{D} = \mathrm{supp}(v) \cap \mathrm{supp}(w)$ of the supports of $v = (v_0, v_1, \ldots, v_{N-1})$ and $w = (w_0, w_1, \ldots, w_{N-1})$ is at most $|\mathcal{D}| = \min\{\mathrm{wt}(v), \mathrm{wt}(w)\} < q-1$. To obtain (6), $v_i + \beta w_i \neq 0$ has to be satisfied for all $i \in \mathcal{D}$. In other words,

$$\beta \neq -\frac{v_i}{w_i} \; \forall \; i \in \mathcal{D} . \tag{7}$$

Equation (7) can always be satisfied choosing a suitable $\beta$, since $|\mathcal{D}| < q-1$, and therefore there exists at least one $\beta \neq 0$, which fulfills (7). ∎

Now, we consider a received word $R = A + E$, where $A$ is a codeword of an IRS code and $E$ is an error matrix. Since the syndromes $\mathcal{S}^{(1)}, \ldots, \mathcal{S}^{(l)}$ calculated by Algorithm 2 only depend on $E$ and not on $A$, we can assume w.l.o.g. the case $A = 0$ to analyze Algorithm 2. In other words, we are able to apply Algorithm 2 directly to the error matrix $E$ instead of the received word $R$.

**Lemma 2** *Assume that $E$ is an $l \times N$ error matrix and $L$ is a non-singular $l \times l$ matrix over $\mathbb{F}_q$. Moreover, let $\widetilde{E} = LE$ be another error matrix, whose rows are linear combinations of the rows of $E$. Then, Algorithm 2 applied to $E$ yields a unique pair $t$, $\Lambda(x)$, if and only if Algorithm 2 applied to $\widetilde{E}$ yields the same unique pair $t$, $\Lambda(x)$.*

*Proof:* Assume that Algorithm 2 is applied to the error matrix

$$E = \begin{pmatrix} e^{(1)} \\ e^{(2)} \\ \vdots \\ e^{(l)} \end{pmatrix} .$$

In its first step, Algorithm 2 computes the Discrete Fourier Transform for every row of $E$:

$$\mathscr{F}(E) = \begin{pmatrix} \mathscr{F}(e^{(1)}) \\ \mathscr{F}(e^{(2)}) \\ \vdots \\ \mathscr{F}(e^{(l)}) \end{pmatrix} = \begin{pmatrix} E_0^{(1)} & E_1^{(1)} & \cdots & E_{N-1}^{(1)} \\ E_0^{(2)} & E_1^{(2)} & \cdots & E_{N-1}^{(2)} \\ \vdots & \vdots & & \vdots \\ E_0^{(l)} & E_1^{(l)} & \cdots & E_{N-1}^{(l)} \end{pmatrix}$$

The last $N - K^{(\ell)}$ elements of the $\ell$th row coincide with the elements of the syndrome sequence $\mathcal{S}^{(\ell)}$. Hence, we formally define the syndrome matrix

$$\begin{aligned}
S(E) &= \begin{pmatrix} \xi & \xi & E_{K^{(1)}}^{(1)} & E_{K^{(1)}+1}^{(1)} & \cdots & E_{N-1}^{(1)} \\ \xi & E_{K^{(2)}}^{(2)} & E_{K^{(2)}+1}^{(2)} & E_{K^{(2)}+2}^{(2)} & \cdots & E_{N-1}^{(2)} \\ \vdots & \vdots & & \vdots & & \\ \xi & \xi & \xi & E_{K^{(l)}}^{(l)} & \cdots & E_{N-1}^{(l)} \end{pmatrix} \\
&= \begin{pmatrix} \xi & \xi & S_0^{(1)} & S_1^{(1)} & \cdots & S_{N-K^{(1)}-1}^{(1)} \\ \xi & S_0^{(2)} & S_1^{(2)} & S_2^{(2)} & \cdots & S_{N-K^{(2)}-1}^{(2)} \\ \vdots & \vdots & & \vdots & & \\ \xi & \xi & \xi & S_0^{(l)} & \cdots & S_{N-1}^{(l)} \end{pmatrix}
\end{aligned}$$

by replacing the first $K^{(\ell)}$ elements of the $\ell$th row by erasure symbols, i.e., by symbols $\xi \notin \mathbb{F}_q$, which fulfill $\alpha + \xi = \xi + \alpha = \xi$, and $\alpha \cdot \xi = \xi \cdot \alpha = \xi$ for an arbitrary element $\alpha \in \mathbb{F}_q$. In other words, $S(E)$ is an $l \times N$ matrix containing the $l$ syndrome sequences aligned to the right, and prepended by erasure symbols. From the linearity of the Discrete Fourier Transform we know that

$$\mathscr{F}(\widetilde{E}) = \mathscr{F}(LE) = L\mathscr{F}(E) ,$$

and consequently also $S(\widetilde{E}) = LS(E)$.

Since (5) is a linear recursion, we know (see e.g. [27]) that if the rows of $S(E)$ satisfy (5) with some pair $t$, $\Lambda(x)$, then the rows of $LS(E)$ also satisfy (5) with the same pair $t$, $\Lambda(x)$. Since $L$ is a non-singular matrix, $E = L^{-1}\widetilde{E}$, and we conclude that $S(E)$ satisfies (5) with the pair $t$, $\Lambda(x)$, if and only if $S(\widetilde{E})$ satisfies (5) with the same pair $t$, $\Lambda(x)$. Thus, if Algorithm 2 has a *unique* solution $t$, $\Lambda(x)$ for $S(E)$, then Algorithm 2 has the same unique solution for $S(\widetilde{E})$, and vice versa. ∎

**Theorem 1 (Guaranteed Correcting Radius)** *Consider an interleaved Reed–Solomon code pursuant to Definition 3. Assume that this code is corrupted by an error matrix $E$ with $t$ non-zero columns. Then, Algorithm 2 always yields a unique and correct solution, i.e., all $t$ column errors can be corrected, as long as $t$ satisfies*

$$t \leq t_{\mathrm{g}} = \left\lfloor \frac{N - \widehat{K}}{2} \right\rfloor , \tag{8}$$

*where*

$$\widehat{K} = \max_{1 \leq \ell \leq l} \left\{ K^{(\ell)} \right\}$$

*is the maximum dimension among the $l$ Reed–Solomon codes $\mathcal{A}^{(1)}, \mathcal{A}^{(2)}, \ldots, \mathcal{A}^{(\ell)}$.*

*Proof:* Assume that a codeword of the interleaved Reed–Solomon code is corrupted by an error matrix $\boldsymbol{E}$ with $t$ non-zero columns. We distinguish two different cases:

1) If $\text{wt}\left(\boldsymbol{e}^{(1)}\right) = \left|\text{supp}\left(\boldsymbol{e}^{(1)}\right)\right| = t$, then the Berlekamp–Massey algorithm applied to the sequence $\mathcal{S}^{(1)}$ yields a pair $t, \Lambda(x)$ which is the shortest length solution of (3), provided that $t$ satisfies (8). Since we consider column errors, the pair $t, \Lambda(x)$ is also a solution of (3) for the remaining sequences $\mathcal{S}^{(2)}, \ldots, \mathcal{S}^{(l)}$. In other words, $t, \Lambda(x)$ is the shortest length solution of (5), which is also obtained by applying Algorithm 1 to the $l$ sequences $\mathcal{S}^{(1)}, \mathcal{S}^{(2)}, \ldots, \mathcal{S}^{(l)}$. Consequently, Algorithm 2 yields the correct solution.

2) Now, we consider the case $\text{wt}\left(\boldsymbol{e}^{(1)}\right) = \left|\text{supp}\left(\boldsymbol{e}^{(1)}\right)\right| < t$, i.e., the case that the column errors only affect $t^{(1)} < t$ symbols in $\boldsymbol{e}^{(1)}$. Since

$$t \leq \left\lfloor \frac{N - \widehat{K}}{2} \right\rfloor < q - 1$$

is always satisfied due to Definition 2, we are able to recursively apply Lemma 1 to obtain $l$ coefficients $\beta_1, \beta_2, \ldots, \beta_l$, $\beta_1 = 1$, such that the linear combination

$$\tilde{\boldsymbol{e}}^{(1)} = \sum_{\ell=1}^{l} \beta_\ell \boldsymbol{e}^{(\ell)}$$

is an error vector with $t$ non-zero coefficients, i.e., we have $\left|\text{supp}\left(\tilde{\boldsymbol{e}}^{(1)}\right)\right| = t$. Hence, we define the non-singular matrix

$$\boldsymbol{L} = \begin{pmatrix} 1 & \beta_2 & \ldots & \beta_l \\ 0 & 1 & \ldots & 0 \\ \vdots & \vdots & \ddots & \vdots \\ 0 & 0 & \ldots & 1 \end{pmatrix},$$

and use it to calculate the error matrix $\widetilde{\boldsymbol{E}} = \boldsymbol{L}\boldsymbol{E}$, which has $t$ non-zero error symbols in its first row. Thus, we know from case 1) that Algorithm 1 synthesizes the correct pair $t, \Lambda(x)$ for $\widetilde{\boldsymbol{E}}$, and that Algorithm 2 consequently yields the correct solution for the matrix $\widetilde{\boldsymbol{E}}$. Due to Lemma 2, we know that Algorithm 2 also yields the correct solution for the error matrix $\boldsymbol{E}$, which proves Theorem 1 for the case 2).

∎

Beyond the guaranteed error correcting radius we may be able to locate the errors, as long as the number of unknowns in (4) is not larger than the number of equations, i.e., as long as

$$t \leq \frac{l}{l+1} \left( N - \frac{1}{l} \sum_{\ell=1}^{l} K^{(\ell)} \right) \qquad (9)$$

is satisfied. Moreover, we are only able to evaluate the errors, if the number of errors does not exceed the number of redundancy symbols in the component code with the largest dimension. This gives rise to the following theorem on the maximum error correcting radius:

**Theorem 2 (Maximum Correcting Radius)** *Consider an interleaved Reed–Solomon code pursuant to Definition 3. Assume that a word of this code is corrupted by $t$ column errors. Then, Algorithm 2 may only find a unique and correct solution, if $t$ satisfies*

$$t \leq t_{\max} = \min\left\{ \frac{l}{l+1}\left(N - \bar{K}\right), N - \widehat{K} \right\}, \qquad (10)$$

*where*

$$\bar{K} = \frac{1}{l} \sum_{\ell=1}^{l} K^{(\ell)}$$

*is the average dimension of the $l$ Reed–Solomon codes.*

*Proof:* To find a unique and correct solution, the linear system of equations (4) with $t$ unknowns and the set of linear recursions (5) of length $t$ must have a unique solution, which means that $\text{rank}(\boldsymbol{S}_l) = t$. This is only possible, if (9) is satisfied, since otherwise the matrix $\boldsymbol{S}_l$ has less than $t$ rows. Hence, in order to be able to locate the errors, $t$ must not be larger than the first argument in (10). Even if the locations of the errors are known, the error values can only be calculated if $t$ is not larger than the number of redundancy symbols in every Reed–Solomon code. Thus, if $t > N - \widehat{K}$, it is not possible to evaluate the errors in at least one Reed–Solomon codeword. Consequently, $t$ must also not be larger than the second argument in (10). ∎

The Theorems 1 and 2 characterize the collaborative decoding approach described by the linear system of equations (4) or equivalently by the set of linear recursions (5). If the number of column errors satisfy (8), i.e., if the number of errors is smaller than half the minimum distance of the Reed–Solomon code with the largest dimension, (4) and hence also (5) have a unique solution, and we are always able to correctly reconstruct the transmitted codeword $\boldsymbol{A}$. If the number of errors lies between the guaranteed error correcting radius specified by (8) and the maximum correcting radius specified by (10), (4) and (5) may still have a unique solution, which enables us to correct errors beyond half the minimum distance of the component codes. In this sense, our collaborative decoding approach yields a *Bounded Distance* (BD) decoder, whose maximum error correction radius lies beyond half the minimum distance. However, as soon as the number of column errors exceeds the limit specified by (8), we are not able any more to guarantee that (4) and hence also (5) has a unique solution. In fact, the solutions of (4) and (5) will be ambiguous with some probability $P_f > 0$, since $\text{rank}(\boldsymbol{S}_l) < t$ for some error patterns.

The possibility to obtain an ambiguous solution is inherent to Bounded Distance decoders which correct errors within a radius beyond half the minimum distance, since some correcting spheres are inevitably overlapping in this case. Basically, three strategies are conceivable to cope with this problem:

1) The decoder selects on single codeword out of all valid decoding results

2) The decoder returns a list of all valid decoding results that is, a list of all codewords which lie inside aسphere with radius $t_{\max}$ around the received word $\boldsymbol{y}$.
3) If the decoding result is not unique, i.e., if there exists more than one valid decoding result, the decoder yields a decoding failure

Since finding all valid solutions is algebraically difficult with our approach and usually requires a high computational effort, the decoders we consider here apply the third strategy: whenever (4) does not have a unique solution, Algorithm 2 may yield a decoding failure (we formalize this statement later in Lemma 5).

## V. DISTRIBUTION OF THE REDUNDANCY

Definition 3 does not restrict the choice of the dimensions of the $l$ Reed–Solomon codes used to construct a heterogeneous IRS code. However, to ensure a good performance under collaborative decoding, we should apply some restrictions on the dimensions $K^{(1)}, \ldots, K^{(l)}$.

From (10) we observe that the maximum error correction radius depends on the average redundancy $N - \bar{K}$ and the minimum redundancy $N - \widehat{K}$ of the components. If the correction radius is limited by the second restriction, our collaborative decoding strategy is not able to recover all errors, even if we are able to locate them. Hence, the redundancy should always be distributed in such a way that we are able to evaluate all errors if we manage to locate them. This is stated by the following theorem:

**Theorem 3 (Redundancy Distribution)** *Consider an interleaved Reed–Solomon code pursuant to Definition 3. For a given code rate $R = \frac{\bar{K}}{N}$, the error correction radius is maximized, if the dimension of the $l$ component codes $\mathcal{A}^{(1)}, \ldots, \mathcal{A}^{(l)}$ are chosen such that they satisfy*

$$\widehat{K} \leq \frac{l}{l+1}\left(\frac{N}{l} + \bar{K}\right). \tag{11}$$

*Proof:* For a given rate $R = \frac{\bar{K}}{N}$, $\frac{l}{l+1}(N - \bar{K})$ is constant in terms of the choice of the dimensions in the codes $\mathcal{A}^{(1)}, \ldots, \mathcal{A}^{(l)}$. Hence, the error correcting radius is maximized for

$$\frac{l}{l+1}(N - \bar{K}) \leq N - \widehat{K}.$$

Solving for $\widehat{K}$ yields

$$\widehat{K} \leq \frac{1}{l+1}N + \frac{l}{l+1}\bar{K} = \frac{l}{l+1}\left(\frac{N}{l} + \bar{K}\right).$$

∎

Clearly, there are not much reasons to consider IRS codes which do not fulfill (11). Therefore, we always assume in the following that IRS codes are designed in accordance to Theorem 3.

By examining (11) we observe that asymptotically, i.e., for $l \to \infty$, the maximum dimension is limited by the average dimension. This means that for IRS code designs with a large $l$, the redundancy should be distributed equally over all component codes. Though, if $l$ is small enough, the redundancy may be distributed unequally, as long as we only care about the maximum error correcting radius, and not about the minimum distance of the IRS code. However, the minimum distance is only optimal in the homogeneous case. This is explained by the fact that homogeneous IRS codes are MDS, while the minimum distance of heterogeneous IRS codes is determined by the Reed–Solomon code with the largest dimension. Nevertheless, heterogeneous IRS codes can be interesting e.g. in the context of *generalized concatenated code* constructions introduced and described by Blokh and Zyablov [13] and Zinoviev [14], for decoding single Reed–Solomon codes beyond half the minimum distance like described in [15], and for several other applications.

## VI. COLLABORATIVE DECODING PERFORMANCE

To evaluate the performance of our decoding strategy, we would like to estimate the error probability $P_e$, and the failure probability $P_f$, which together result in the probability $P_w$ for obtaining a wrong decoding result. We are particularly interested in the failure probability $P_f(t)$ in the range $t_g < t \leq t_{\max}$, since the proposed decoding strategy certainly makes only sense if $P_f(t)$ is small enough, such that the decoder only fails in a very few cases as long as $t$ is below the maximum error correcting radius $t_{\max}$.

### A. Maximum Likelihood Certificate Property

Before we derive upper bounds on $P_f$ and $P_e$, we consider another interesting property of our collaborative decoding strategy: we show that Algorithm 2 exhibits the *Maximum Likelihood (ML) certificate* property. This means that whenever the decoder does not fail, then it yields the ML solution, i.e., the solution with the minimum Hamming distance to the received word. This property will help us later, to overbound $P_e$.

**Definition 5 (ML Certificate)**
*Consider a code $\mathcal{C}$, and assume that the word $\boldsymbol{y}$ is received, when a codeword $\boldsymbol{c} \in \mathcal{C}$ is transmitted over a memoryless noisy channel. Moreover, consider a decoding algorithm which either decodes a codeword $\hat{\boldsymbol{c}} \in \mathcal{C}$ or yields a decoding failure. We say that the decoding algorithm exhibits the ML certificate property, if whenever the decoder decides on a codeword $\hat{\boldsymbol{c}} \in \mathcal{C}$, there does not exist another codeword $\boldsymbol{c}' \in \mathcal{C}$, which has a smaller Hamming distance to $\boldsymbol{y}$ than $\hat{\boldsymbol{c}}$.*

To show that Algorithm 2 exhibits the ML certificate property, we state the following lemma, which immediately follows from the definitions of the error locator polynomial and the syndromes in Section III:

**Lemma 3** *Consider a codeword $\boldsymbol{A} \in \mathcal{A}$ of an IRS code according to Definition 3. Assume that this word is corrupted by an error matrix $\boldsymbol{E}$ with $t$ non-zero columns at the positions $j_1, j_2, \ldots, j_t$. Hence, we observe a matrix $\boldsymbol{R} = \boldsymbol{A} + \boldsymbol{E}$ with $l$ rows, from which we are able to compute the syndromes $\mathcal{S}^{(1)}, \mathcal{S}^{(2)}, \ldots, \mathcal{S}^{(l)}$. Then, the error locator polynomial $\Lambda(x) = \prod_{i=1}^{t}\left(1 - \alpha^{-j_i}x\right)$ is a t-valid polynomial, which* *is a solution of the system of equations (4) with $t$ unknowns, and of the set of linear recursions (5) of length $t$.*

Lemma 3 is helpful for proving the following theorem:

**Theorem 4 (ML Certificate of Algorithm 2)** *The decoder specified by Algorithm 2 exhibits the ML certificate property.*

*Proof:* If Algorithm 2 does not fail while decoding $\boldsymbol{R}$, it yields a pair $t$, $\Lambda(x)$, where $\Lambda(x)$ is a $t$-valid polynomial satisfying the system of equations (4) with $t$ unknowns, and the set of linear recursions (5) of length $t$. From $\Lambda(x)$, the decoder computes an error word $\boldsymbol{E}$ with $t$ non-zero columns, and a codeword $\boldsymbol{A} = \boldsymbol{R} - \boldsymbol{E}$ which differs in $t$ columns from $\boldsymbol{R}$. Hence, the extension field representations $\mathbf{a}$, and $\mathbf{r}$ of the matrices $\boldsymbol{A}$ and $\boldsymbol{R}$ fulfill $d(\mathbf{a}, \mathbf{r}) = t$.

Now, we assume that there exists a codeword $\boldsymbol{A}'$ whose extension field representation $\mathbf{a}'$ has a smaller Hamming distance to $\mathbf{r}$ than $\mathbf{a}$, i.e.,

$$d(\mathbf{a}', \mathbf{r}) = t' < t.$$

Applying Lemma 3 to $\boldsymbol{R}$ and $\boldsymbol{A}'$ yields that there exists a $t'$-valid solution $\Lambda'(x)$ of the system of equations (4) with $t'$ unknowns and the set of linear recursions (5) of length $t'$, where $t' < t$. However, since Algorithm 1 always finds a solution of (5) with the smallest $t$, this yields a contradiction to the assumption that $\text{dist}(\mathbf{a}', \mathbf{r}) < \text{dist}(\mathbf{a}, \mathbf{r})$. Hence, if Algorithm 2 finds a solution, this solution corresponds to the codeword with the smallest Hamming distance to $\boldsymbol{R}$. ∎

### B. Error Probability

Now, we derive an upper bound on $P_e$ for the general case that a linear block code is decoded by a BD decoder which exhibits the ML certificate property. For this purpose, we assume that the transmitted word is corrupted by $t$ errors, and that the BD decoder is able to correct errors up to the radius $t_{\max}$. We consider a pair of codewords, and count the number of received vectors which may be decoded into wrong codewords. Then, we use a union bounding technique to overbound the error probability $P_e(t)$. The bound obtained in this way is directly applicable to IRS codes by interpreting the IRS codewords as vectors over the extension field $\mathbb{F}_{q^l}$.

**Lemma 4** *Consider two concentric spheres $s_1$ and $s_2$ in the Hamming space $\mathbb{F}_q^N$. Assume that $s_1$ has radius $r_1$, and $s_2$ has radius $r_2$. Fix an arbitrary point $\boldsymbol{v}$ on the surface of $s_1$, and let $U(q, r_2, r_1, \rho)$ be the number of points on the surface of $s_2$ with distance $\rho$ to $\boldsymbol{v}$. Then, $U(q, r_2, r_1, \rho)$ is calculated by*

$$U(q, r_2, r_1, \rho) = \sum_{i=\lceil \frac{r_1+r_2-\rho}{2} \rceil}^{r_1+r_2-\rho} \binom{r_1}{i} \binom{i}{\rho - (r_1+r_2) + 2i} \binom{N-r_1}{r_2-i} \cdot (q-2)^{\rho-(r_1+r_2)+2i} (q-1)^{r_2-i}.$$

*Proof:* For proving Lemma 4, we assume w.l.o.g. that the spheres $s_1$ and $s_2$ are centered around the origin. Then, we consider a vector $\boldsymbol{v}$ on $s_1$. Since $s_1$ is centered around the origin, $\text{wt}(\boldsymbol{v}) = r_1$, and hence $\boldsymbol{v}$ has exactly $r_1$ non-zero components. Now, we consider another vector $\boldsymbol{w}$ on the surface of $s_2$, i.e., $\text{wt}(\boldsymbol{w}) = r_2$, and assume that $\boldsymbol{v}$ and $\boldsymbol{w}$ are overlapping in exactly $i$ non-zero coordinates. The Hamming distance $d(\boldsymbol{v}, \boldsymbol{w})$ is calculated by

$$d(\boldsymbol{v}, \boldsymbol{w}) = (r_1 - i) + (r_2 - i) + \delta(i),$$

where $\delta(i)$ is number of differing symbols in the overlapping coordinates. Then, we count how many vectors $\boldsymbol{w}$ exist, for which $d(\boldsymbol{v}, \boldsymbol{w}) = \rho$, or equivalently

$$\delta(i) = \rho - (r_1 + r_2) + 2i. \tag{12}$$

Hence, we have

$$\binom{i}{\rho - (r_1 + r_2) + 2i}(q-2)^{\rho - (r_1+r_2) + 2i}$$

possibilities to choose non-zero symbols in the overlapping coordinates which are different in the two vectors, and

$$\binom{N - r_1}{r_2 - i}(q-1)^{r_2 - i}$$

non-zero symbols in the non-overlapping coordinates of $\boldsymbol{w}$, in order to obtain a vector with Hamming distance $\rho$ to $\boldsymbol{v}$. Consequently, to calculate $U(q, r_2, r_1, \rho)$, we have to sum up the product

$$\binom{i}{\rho - (r_1+r_2)+2i} \binom{N-r_1}{r_2-i} \cdot \\ \cdot (q-2)^{\rho-(r_1+r_2)+2i}(q-1)^{r_2-i}$$

for all possible numbers of overlapping coordinates. To obtain the summation limits, we observe that $\delta(i)$ is limited by $0 \leq \delta(i) \leq i$. By inserting (12), we obtain

$$(r_1 + r_2 - \rho)/2 \leq i \leq r_1 + r_2 - \rho.$$

Hence we have to sum up from $\lceil \frac{r_1+r_2-\rho}{2} \rceil$ to $r_1 + r_2 - \rho$. Consequently, $U(q, r_2, r_1, \rho)$ is calculated by the sum

$$\sum_{i=\lceil \frac{r_1+r_2-\rho}{2} \rceil}^{r_1+r_2-\rho} \binom{r_1}{i}\binom{i}{\rho - (r_1+r_2)+2i}\binom{N-r_1}{r_2-i} \cdot \\ \cdot (q-2)^{\rho-(r_1+r_2)+2i}(q-1)^{r_2-i}.$$

This proves Lemma 4. ∎

Theorem 4 and Lemma 4 enable us to prove the following theorem, which overbounds the error probability $P_e(t)$:

**Theorem 5 (Error Probability)** *Let $\mathcal{C}(Q; N, K, D)$ be a linear block code of length $N$, dimension $K$, and minimum distance $D$ over the field $\mathbb{F}_Q$, decoded by a BD decoder which exhibits the ML certificate property. Assume that the decoding radius of this decoder is $t_{\max}$, and that it decodes a received word, which is corrupted by $t$ errors. Then, the probability for a decoding error is overbounded by*

$$P_e(t) \leq \overline{P}_e(t) = \frac{\sum_{w=D}^{t+t_{\max}} A_w \sum_{\rho=0}^{\min\{t, t_{\max}\}} \cdot U(Q, t, w, \rho)}{\binom{N}{t}(Q-1)^t}, \tag{13}$$

where $A_w$ describes the weight distribution of the code, i.e., $A_w$ is the number of codewords of weight $w$.

*Proof:* Consider two codewords $\boldsymbol{c} \in \mathcal{C}$ and $\boldsymbol{c}' \in \mathcal{C}$ with Hamming distance $\mathrm{dist}\,(\boldsymbol{c}, \boldsymbol{c}') = w$ to each other. Assume that the codeword $\boldsymbol{c}$ is corrupted by $t$ errors, and that the word $\boldsymbol{y}$ is observed at the output of the channel. For this arrangement, we overbound the pairwise error probability $P(\boldsymbol{c} \to \boldsymbol{c}')$, i.e, the probability that $\boldsymbol{c}'$ is decoded under the condition that $\boldsymbol{c}$ has been transmitted. To do this, we consider a sphere $s$ of radius $t$ centered at $\boldsymbol{c}$. We count all points on the surface of $s$, which lie inside a sphere $s'$ of radius $t_{\max}$ around $\boldsymbol{c}'$, since they may be decoded into $\boldsymbol{c}'$ (see Fig. 2). However, we only count the points on $s'$, which have a maximum distance of $t$ to $\boldsymbol{c}'$, since otherwise they are closer to $\boldsymbol{c}$ than to $\boldsymbol{c}'$, and due to the ML certificate property we know that they are correctly decoded into the codeword $\boldsymbol{c}$.

In other words, we count all points on $s$, whose distance $\rho$ to $\boldsymbol{c}'$ is in the range $0 \le \rho \le \min\{t, t_{\max}\}$. For this purpose, we consider the two concentric spheres $s_1$ and $s_2$ around $\boldsymbol{c}$. We select the sphere $s_1$ to have radius $w$, and the sphere $s_2$ to have radius $t$. This means that $\boldsymbol{c}'$ is a point on $s_1$, and $s_2$ coincides with the sphere $s$. Hence we are able to apply Lemma 4 to count the number of points $N_p$ on the surface of $s_2$ with distance $\rho$ to $\boldsymbol{c}'$ on $s_1$, where $0 \le \rho \le \min\{t, t_{\max}\}$. This is illustrated by Fig. 2.

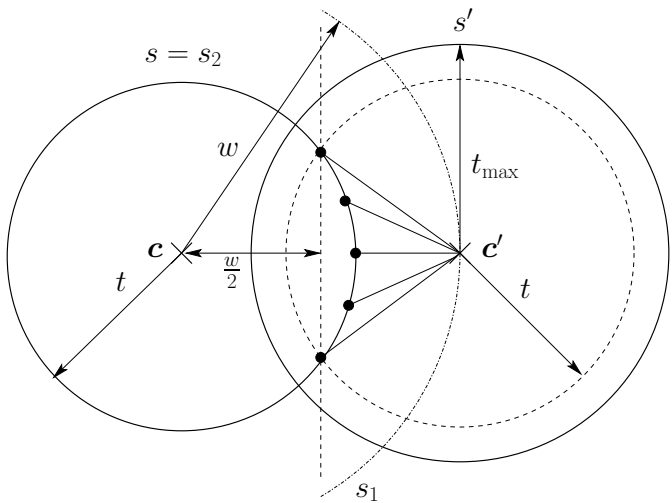

Fig. 2. Number of points which may contribute to $P(\boldsymbol{c} \to \boldsymbol{c}')$

In this way, we obtain
$$N_p = \sum_{\rho=0}^{\min\{t, t_{\max}\}} U(Q, t, w, \rho)\,.$$

To overbound the total number $N_t$ of points which are decoded into wrong codewords, we use a union bounding technique and consider all codewords within distance $D \le w \le t + t_{\max}$ to $\boldsymbol{c}$, since the sphere $s$ does not overlap with the correcting spheres of all other codewords. As there are $A_w$ codewords with distance $w$ to $\boldsymbol{c}$, we obtain

$$N_t \le \sum_{w=D}^{t+t_{\max}} A_w N_p\,. \tag{14}$$

To calculate $P_e(t)$, we divide $N_t$ by the number $S_n(t)$ of points on the sphere $s$ with radius $t$, which is calculated by

$$S_n(t) = \binom{N}{t}(Q-1)^t\,.$$

Since $N_t$ is overbounded by (14), we obtain the statement of Theorem 5. ∎

Note that if we set $t_{\max} = t_{\mathrm{g}}$ in Theorem 5, (13) coincides with the error probability for classical BMD decoding as described in [24]. The expression is actually exact in this case, since the spheres around the surrounding codewords do not overlap in the case $t_{\max} = t_{\mathrm{g}}$.

Theorem 5 basically holds for arbitrary linear block codes over the field $\mathbb{F}_q$, which are decoded up to the maximum error correcting radius $t_{\max}$ by a decoder exhibiting the ML certificate property. To apply this theorem to IRS codes, we interpret them as codes over the extension field $\mathbb{F}_{q^l}$, and set $Q = q^l$. Due to Theorem 4, Algorithm 2 exhibits the ML certificate property, and hence (13) provides us with an upper bound on the error probability $P_e(t)$. However, to calculate (13), we require knowledge about the weight distribution of the code. A homogeneous IRS code can be interpreted as an MDS code over the extension field $\mathbb{F}_{q^l}$. Since the weight distribution of MDS codes is well–known (see e.g. [28]), the weights $A_w$ can be calculated for this case by

$$A_w = \binom{N}{w}(q^l - 1)\sum_{i=0}^{w-D}(-1)^i\binom{w-1}{i}q^{l(w-D-i)}\,,$$

where $D = N - K + 1$. Unfortunately, heterogeneous IRS codes are usually not MDS. Hence $A_w$ can generally not be obtained in such a simple way. However, even if the weight distribution is not known for a heterogeneous IRS code, we are able to overbound $P_e(t)$ on the basis of (13), by replacing $A_w$ by $\widehat{A}_w$, where $\widehat{A}_w$ is such that $\widehat{A}_w \ge A_w \,\forall\, w = 0, \ldots, N$.

*C. Failure Probability*

Collaborative decoding of IRS codes provides us with a method of decoding errors beyond half the minimum distance, if the errors affect the columns of the interleaved scheme like depicted in Fig. 1. If the number of errors is smaller than the guaranteed correcting radius (8), we know from Theorem 1 that we are always able to correct these errors. Contrariwise, if the number of errors exceeds the maximum correcting radius (10), the decoder will always fail or take an erroneous decision.

In order to analyze the probability $P_f(t)$ in the range $t_{\mathrm{g}} < t \le t_{\max}$, we assume that for each column of the interleaved Reed–Solomon code, each error pattern occurs equiprobable. More precisely, we assume, that the burst errors

$$\boldsymbol{e}_j = \begin{pmatrix} e_j^{(1)} \\ \vdots \\ e_j^{(l)} \end{pmatrix}$$

are random vectors, uniformly distributed over $\mathbb{F}_q^l \setminus \{\boldsymbol{0}\}$. Under this assumption, bounds on $P_f$ have been derived in [4], and [5] for homogeneous IRS codes. However, since these bounds are rather weak or do not depend on the error weight $t$, they generally do not yield a good estimation of the decoding performance.

To obtain a bound for heterogeneous IRS codes, we generalize a bound from [10] for homogeneous IRS codes, which allows for estimating the failure probability $P_f(t)$ in dependence of $t$. For proving this generalized bound, we use similar but simpler techniques, which have been applied in [2] to bound the failure probability for *folded Reed–Solomon codes*. These techniques are based upon analyzing the linear system of equations (4), and overbounding the number of cases, in which the system of equations (4) with $t$ unknowns yields multiple solutions. To obtain an upper bound on the failure probability of Algorithm 2 from this analysis, we have to show, that whenever Algorithm 2 yields a decoding failure, then there exist multiple solutions for (4).

**Lemma 5** *Consider a codeword $\boldsymbol{A} \in \boldsymbol{\mathcal{A}}$ of an IRS code pursuant to Definition 3. Assume that this word is corrupted by an error matrix $\boldsymbol{E}$ with $t$ non-zero columns, and that Algorithm 2 yields a decoding failure. Then, the linear system of equations (4) with $t$ unknowns has multiple solutions.*

*Proof:* Assume that a codeword $\boldsymbol{A} \in \boldsymbol{\mathcal{A}}$ has been transmitted and $\boldsymbol{Y} = \boldsymbol{A} + \boldsymbol{E}$ is received, where $\boldsymbol{E}$ has $t$ non-zero columns. Algorithm 2 only yields a decoding failure, if Algorithm 1 computes a pair $t', \Lambda'(x)$, such that $\Lambda'(x)$ is not $t'$-valid.

First we show that $t' \leq t$. Applying Lemma 3 to $\boldsymbol{Y}$ and $\boldsymbol{A}$ yields that there exists a $t$-valid polynomial $\Lambda(x)$, which is a solution of (5). Since Algorithm 1 always yields the smallest length shift-register which is a solution of (5), we know that $t' \leq t$.

Second, $\Lambda'(x)$ is a solution of the set of linear recursions (5) of length $t'$, and hence also for the system of equations (4) with $t'$ unknowns. However, if $\Lambda'(x)$ is a solution for the set of linear recursions (5) of length $t'$, it is also a solution for length $t \geq t'$, since if $\Lambda'(x)$ fulfills (5) for $i \geq t'$, it obviously also fulfills it for $i \geq t \geq t'$. Hence, $\Lambda'(x)$ is not only a solution of the system of equations (4) with $t'$ unknowns and the set of linear recursions (5) of length $t'$, but also for the system (4) with $t$ unknowns, and the set of recursions (5) of length $t$. Consequently, the system of equations (4) with $t$ unknowns has at least the two solutions $\Lambda(x)$ and $\Lambda'(x)$. This proves the statement of Lemma 5. ∎

Note that we only show that if Algorithm 2 yields a decoding failure, then (4) has multiple solutions. Contrariwise this does not necessarily mean that Algorithm 2 always fails if (4) has multiple solutions. However, since we are only interested in overbounding the failure probability $P_f$, Lemma 5 is sufficient for our purposes, since the number of events for which (4) has multiple solutions is an upper bound for the number of events in which Algorithm 2 fails.

Hence, to overbound the probability $P_f$ that Algorithm 2 fails, we consider the following lemma:

**Lemma 6** *Let $\mathcal{K}^{(1)}, \mathcal{K}^{(2)}, \ldots, \mathcal{K}^{(l)}$ be $l$ $q$-ary linear codes of length $\omega$, and let the dimension of the code $\mathcal{K}^{(\ell)}$ be $\omega - \varrho^{(\ell)}$.*

*Further, let*

$$\boldsymbol{W} = \begin{pmatrix} \boldsymbol{w}^{(1)} \\ \vdots \\ \boldsymbol{w}^{(l)} \end{pmatrix} = (\boldsymbol{w}_0, \ldots, \boldsymbol{w}_{\omega-1})$$

*be a $l \times \omega$ matrix, such that $\boldsymbol{w}_j \neq \boldsymbol{0}, \ \forall j = 0, \ldots, \omega - 1$, i.e., that $\boldsymbol{W}$ does not have any all-zero column. Furthermore, assume that all columns of $\boldsymbol{W}$ are uniformly distributed over all non-zero vectors of length $l$. Then, the probability $P_\omega$ that*

$$\boldsymbol{w}^{(\ell)} \in \mathcal{K}^{(\ell)}, \ \forall \ell = 1, \ldots, l \tag{15}$$

*is overbounded by*

$$P_\omega \leq \frac{q^{l\omega}}{(q^l - 1)^\omega} \cdot q^{-\sum_{\ell=1}^l \varrho^{(\ell)}} . \tag{16}$$

*Proof:* Let $\mathcal{L}$ be the set of all $l \times \omega$ matrices whose rows fulfill Equation (15). Further, let $\mathcal{S}_\omega$ be the set of all $l \times \omega$ matrices with elements from $\mathbb{F}_q$, and let the subset $\mathcal{S}_\omega^v \subset \mathcal{S}_\omega$ be the set of matrices without any non-zero column. Then, the probability $P_\omega$ that a matrix $\boldsymbol{W}$ without any all-zero column fulfills (15) can be calculated by

$$P_\omega = \frac{|\mathcal{L} \cap \mathcal{S}_\omega^v|}{|\mathcal{S}_\omega^v|} \leq \frac{|\mathcal{L}|}{|\mathcal{S}_\omega^v|} .$$

The cardinality $|\mathcal{L}|$ is obtained by

$$|\mathcal{L}| = \prod_{\ell=1}^l \left| \mathcal{K}^{(\ell)} \right| = q^{l\omega - \sum_{\ell=1}^l \varrho^{(\ell)}} ,$$

and the cardinality $|\mathcal{S}_\omega^v|$ is calculated by

$$|\mathcal{S}_\omega^v| = \left( q^l - 1 \right)^\omega .$$

Consequently, $P_\omega$ is overbounded by

$$P_\omega \leq \frac{q^{l\omega}}{(q^l - 1)^\omega} \cdot q^{-\sum_{\ell=1}^l \varrho^{(\ell)}} .$$

∎

Lemma 6 enables us, to state and prove the following theorem:

**Theorem 6 (Failure Probability)** *Consider an interleaved Reed–Solomon code pursuant to Definition 3, which is decoded by Algorithm 2. Assume that the codes $\mathcal{A}^{(1)}, \ldots, \mathcal{A}^{(l)}$ are chosen such that (11) is satisfied. Furthermore assume that $\boldsymbol{A}$ is corrupted by $t$ column errors, where each column vector is an independent random vector uniformly distributed over $\mathbb{F}_q^l \setminus \{\boldsymbol{0}\}$. Then, the probability for a decoding failure is overbounded by*

$$P_f(t) \leq \overline{P}_f(t) = \left( \frac{q^l - \frac{1}{q}}{q^l - 1} \right)^t \cdot \frac{q^{-(l+1)(t_{\max} - t)}}{q - 1} , \tag{17}$$

*where $t_{\max} = \frac{l}{l+1}(N - \bar{K})$ is the maximum error correcting radius.*

*Proof:* According to Lemma 5, the failure probability of Algorithm 2 can be overbounded by considering the cases,

in which the system of equations (4) with $t$ unknowns has multiple solutions. We have such a case whenever $\mathrm{rank}\,(\boldsymbol{S}_l) < t$, i.e., whenever there exists a column vector $\boldsymbol{u} \neq \boldsymbol{0}$, such that $\boldsymbol{S}_l \cdot \boldsymbol{u} = \boldsymbol{0}$. Equivalently we can say that (4) cannot have a unique solution, if

$$\exists\, \boldsymbol{u} \neq \boldsymbol{0}\,:\, \boldsymbol{S}^{(\ell)} \cdot \boldsymbol{u} = \boldsymbol{0}\ \forall\, \ell = 1, \ldots, l\,. \tag{18}$$

Since the syndrome matrices $\boldsymbol{S}^{(1)}, \boldsymbol{S}^{(2)}, \ldots, \boldsymbol{S}^{(l)}$ directly depend on the error matrix $\boldsymbol{E}$, we are able to express the failure probability $P_f(t)$ in a general way by

$$P_f(t) = \frac{\text{number of matrices } \boldsymbol{E}_t \text{ satisfying (18)}}{\text{total number of matrices } \boldsymbol{E}_t}\,,$$

where $\boldsymbol{E}_t$ denotes an error matrix with exactly $t$ non-zero columns. Now, we consider matrices with non-zero columns at fixed indices $j_1, j_2, \ldots, j_t$. More precisely, for a fixed set $\{j_1, j_2, \ldots, j_t\}$ of $t$ indices, we consider the ensemble $\mathcal{E}_t(j_1, \ldots, j_t)$ of matrices, in which every column with index $j \in \{j_1, j_2, \ldots, j_t\}$ is an independent random vector uniformly distributed over $\mathbb{F}_q^l \setminus \{\boldsymbol{0}\}$, and all other columns are zero vectors. Then, the probability that (18) is satisfied for matrices $\boldsymbol{E}_t$ from the ensemble $\mathcal{E}_t(j_1, \ldots, j_t)$ is calculated by

$$P_f(j_1, \ldots, j_t) = \frac{|\{\boldsymbol{E} \in \mathcal{E}_t(j_1, \ldots, j_t) : \boldsymbol{E} \text{ satisfies (18)}\}|}{|\mathcal{E}_t(j_1, \ldots, j_t)|}\,.$$

We will now derive an upper bound on $P_f(j_1, \ldots, j_t)$, which does not depend on the selection of the indices $j_1, j_2, \ldots, j_t$, but only on the number of erroneous columns $t$. Hence, this bound will directly provide us with the upper bound on $P_f(t)$, in which we are actually interested in.

For calculating $P_f(j_1, \ldots, j_t)$, let the number of rows in $\boldsymbol{S}^{(\ell)}$ be denoted by $\varrho^{(\ell)}$, i.e., $\varrho^{(\ell)} = N - K^{(\ell)} - t$. It is known (cf. e.g. [24]) that a syndrome matrix $\boldsymbol{S}^{(\ell)}$ can be decomposed into

$$\boldsymbol{S}^{(\ell)} = \boldsymbol{H}^{(\ell)} \cdot \boldsymbol{F}^{(\ell)} \cdot \boldsymbol{D} \cdot \boldsymbol{V}\,.$$

At this, the matrix,

$$\boldsymbol{V} = \begin{pmatrix} 1 & \alpha^{j_1} & \alpha^{2j_1} & \ldots & \alpha^{(t-1)j_1} \\ 1 & \alpha^{j_2} & \alpha^{2j_2} & \ldots & \alpha^{(t-1)j_2} \\ \vdots & \vdots & & \vdots & \\ 1 & \alpha^{j_t} & \alpha^{2j_t} & \ldots & \alpha^{(t-1)j_t} \end{pmatrix}$$

is a $t \times t$ *Vandermonde matrix*, the matrices

$$\boldsymbol{D} = \mathrm{diag}\left(\alpha^{j_1}, \alpha^{j_2}, \ldots, \alpha^{j_t}\right)$$

and

$$\boldsymbol{F}^{(\ell)} = \mathrm{diag}\left(e_{j_1}^{(\ell)}, e_{j_2}^{(\ell)}, \ldots, e_{j_t}^{(\ell)}\right)$$

are $t \times t$ diagonal matrices, and the matrix

$$\boldsymbol{H}^{(\ell)} = \begin{pmatrix} 1 & 1 & \ldots & 1 \\ \alpha^{j_1} & \alpha^{j_2} & \ldots & \alpha^{j_t} \\ \alpha^{2j_1} & \alpha^{2j_2} & \ldots & \alpha^{2j_t} \\ \vdots & \vdots & & \vdots \\ \alpha^{(\varrho^{(\ell)}-1)j_1} & \alpha^{(\varrho^{(\ell)}-1)j_2} & \ldots & \alpha^{(\varrho^{(\ell)}-1)j_t} \end{pmatrix}$$

is a $\varrho^{(\ell)} \times t$ matrix consisting of $\varrho^{(\ell)}$ rows of a transposed Vandermonde matrix. Hence, $\boldsymbol{H}^{(\ell)}$ represents a parity-check matrix of a (shortened) Reed–Solomon code of length $t$ and dimension $t - \varrho^{(\ell)}$, which we denote by $\mathcal{K}^{(\ell)}$. We observe that the matrices $\boldsymbol{V}$ and $\boldsymbol{D}$ both have full rank. Therefore, the product $\boldsymbol{v} = \boldsymbol{D} \cdot \boldsymbol{V} \cdot \boldsymbol{u}$ defines a one-to-one mapping $\boldsymbol{u} \mapsto \boldsymbol{v}$, such that $\boldsymbol{0} \mapsto \boldsymbol{0}$. Consequently, the statement

$$\exists\, \boldsymbol{v} \neq \boldsymbol{0}\,:\, \boldsymbol{H}^{(\ell)} \cdot \boldsymbol{F}^{(\ell)} \cdot \boldsymbol{v} = \boldsymbol{0}\ \forall\, \ell = 1, \ldots, l \tag{19}$$

is equivalent to Equation (18). With $\boldsymbol{w}^{(\ell)} = \left(\boldsymbol{F}^{(\ell)} \cdot \boldsymbol{v}\right)^\mathsf{T}$, and the fact that $\boldsymbol{H}^{(\ell)}$ is a parity-check matrix of the code $\mathcal{K}^{(\ell)}$, we can state another equivalent condition for a decoding failure:

$$\exists\, \boldsymbol{v} \neq \boldsymbol{0}\,:\, \boldsymbol{w}^{(\ell)} \in \mathcal{K}^{(\ell)}\ \forall\, \ell = 1, \ldots, l\,.$$

Assume that we have a vector $\boldsymbol{v}$ with Hamming weight $\mathrm{wt}(\boldsymbol{v}) = \omega$. Then, the vectors $\boldsymbol{w}^{(\ell)}$, $\ell = 1, \ldots, l$, have at most $\omega$ non-zero components. Now consider the matrix

$$\boldsymbol{W} = \begin{pmatrix} \boldsymbol{w}^{(1)} \\ \vdots \\ \boldsymbol{w}^{(l)} \end{pmatrix}\,.$$

Since we know that all vectors $\boldsymbol{e}_{j_i} = \left(e_{j_i}^{(1)}, \ldots, e_{j_i}^{(l)}\right)^\mathsf{T}$, $i = 1, \ldots, t$, are non-zero, and that all non-zero error patterns are distributed uniformly, we also know that $\boldsymbol{W}$ contains exactly $\omega$ non-zero columns uniformly distributed over all non-zero vectors in $\mathbb{F}_q^l$. Assume that the non-zero columns in $\boldsymbol{W}$ are located at the indices $i_1, i_2, \ldots, i_\omega$, let $\boldsymbol{W}_\omega$ be a $l \times \omega$ matrix consisting of the non-zero columns of $\boldsymbol{W}$, and let $\boldsymbol{H}_\omega^{(\ell)}$ be obtained from $\boldsymbol{H}^{(\ell)}$ by removing all columns whose indices are not in the set $\{i_1, i_2, \ldots, i_\omega\}$. Furthermore, denote by $\mathcal{K}_\omega^{(\ell)}$ the code defined by the $\omega \times \varrho^{(\ell)}$ parity-check matrix $\boldsymbol{H}_\omega^{(\ell)}$. Then, the statements $\boldsymbol{H}^{(\ell)} \cdot \boldsymbol{W}^\mathsf{T} = \boldsymbol{0}$, and $\boldsymbol{H}_\omega^{(\ell)} \cdot \boldsymbol{W}_\omega^\mathsf{T} = \boldsymbol{0}$ are equivalent. Consequently, we can apply Lemma 6 on $\mathcal{K}_\omega^{(\ell)}$, and $\boldsymbol{W}_\omega$ to overbound the probability $P_\omega$ that a fixed vector $\boldsymbol{v}$ of weight $\omega$ satisfies

$$\boldsymbol{H}^{(\ell)} \cdot \boldsymbol{F}^{(\ell)} \cdot \boldsymbol{v} = 0\ \forall\, \ell = 1, \ldots, l\,. \tag{20}$$

We observe that the probability $P(\boldsymbol{v})$ for a vector $\boldsymbol{v}$ to fulfill (20) is independent of the positions of the non-zero symbols in $\boldsymbol{v}$, but only depends on the weight $\mathrm{wt}(\boldsymbol{v}) = \omega$, i.e.,

$$P(\boldsymbol{v} : \mathrm{wt}(\boldsymbol{v}) = \omega) = P_\omega\,.$$

Hence, the probability $P_f(j_1, \ldots, j_t)$ that (18) or equivalently (19) is satisfied can be overbounded using a union bounding technique, by summing up over all non-zero vectors $\boldsymbol{v}$:

$$P_f(j_1, \ldots, j_t) \leq \sum_{\boldsymbol{v} \in \mathbb{F}_q^l \setminus \{\boldsymbol{0}\}} P(\boldsymbol{v}) = \sum_{\omega=1}^t \sum_{\{\boldsymbol{v}:\mathrm{wt}(\boldsymbol{v})=\omega\}} P_\omega\,. \tag{21}$$

Since the right side of (21) is independent of the indices $j_1, j_2, \ldots, j_t$ but only depends on $t$, we see that (21) is also an upper bound on the failure probability $P_f(t)$.

To improve (21), we should take care about the following fact: if a vector $\boldsymbol{v}$ fulfills (20), a vector $\boldsymbol{v}' = \alpha \boldsymbol{v}$ also fulfills (20) for all $\alpha \in \mathbb{F}_q \setminus \{0\}$. Therefore, we call $\boldsymbol{v}$ and $\alpha \boldsymbol{v}$ *equivalent vectors*. Since there are $q - 1$ different non-zero elements in $\mathbb{F}_q$, there exist $q - 1$ equivalent vectors for each non-zero vector over $\mathbb{F}_q$. Thus, the number $M_\omega$ of non-equivalent vectors of length $t$ with a certain weight $\omega$ is calculated by

$$M_\omega = \binom{t}{\omega}\frac{(q-1)^\omega}{q-1} = \binom{t}{\omega}(q-1)^{\omega-1}. \quad (22)$$

Hence, to obtain a better upper bound on $P_f(t)$, we can multiply the probabilities $P_\omega$ bounded by (16) by the number of non-equivalent words of weight $\omega$ calculated by (22), and sum up over all weights $1 \leq \omega \leq t$. In this way we obtain

$$P_f(t) \leq \sum_{\omega=1}^{t} P_\omega \cdot M_\omega \leq$$

$$\leq \frac{q^{-\sum_{\ell=1}^{l}\varrho^{(\ell)}}}{q-1}\sum_{\omega=0}^{t}\binom{t}{\omega}\left(\frac{(q-1)q^l}{q^l-1}\right)^\omega =$$

$$= \left(\frac{q^{l+1}-1}{q^l-1}\right)^t \cdot \frac{q^{-\sum_{\ell=1}^{l}\varrho^{(\ell)}}}{q-1} =$$

$$= \left(\frac{q^l - \frac{1}{q}}{q^l-1}\right)^t \cdot \frac{q^{-l(N-\bar{K})+(l+1)t}}{q-1} =$$

$$= \left(\frac{q^l - \frac{1}{q}}{q^l-1}\right)^t \cdot \frac{q^{-(l+1)(t_{\max}-t)}}{q-1}.$$

This proves Theorem 6. ∎

For the case of homogeneous IRS codes, the bound (17) was mentioned the first time in [9] without proof. A similar but somewhat weaker bound is presented in [6] for decoding homogeneous *interleaved Algebraic Geometry* (AG) codes. Applied to homogeneous IRS codes, this bound basically differs from (17) by the factor $\frac{1}{q-1}$.

### D. Comparison of $P_f$ and $P_e$

The decoding performance of an IRS scheme is mainly influenced by the probabilities $P_f(t)$ and $P_e(t)$ in the range $t_{\rm g} \leq t \leq t_{\max}$. To get an impression of these probabilities, we consider a homogeneous IRS code composed of three codewords of the code $\mathcal{A} = \mathcal{RS}(2^8; 255, 223, 33)$. This Reed–Solomon code is used in several practical applications, and guarantees to decode all error patterns with a weight up to 16. The resulting IRS code $\mathcal{A}$ can be considered as MDS code of length 255, dimension 223, and minimum distance 33 over the field $\mathbb{F}_{(2^8)^3}$. According to Theorem 2, our collaborative decoding strategy is able to correct up to 24 errors. Hence, we calculate the upper bound on $P_f(t)$ pursuant to (17), and the upper bound on $P_e(t)$ pursuant to (13) in the interval $17 \leq t \leq 24$ to figure out, in how many cases the decoder fails or yields an erroneous result in the worst case. The results are depicted by Figure 3. We observe that both probabilities decrease exponentially with decreasing $t$, and that the the error probability $P_e(t)$ is smaller than the failure probability $P_f(t)$ by several orders of magnitude in the complete range. Hence, the decoder performance is usually dominated by the failure probability $P_f$, and not by the error probability $P_e$.

### E. Overall Word Error Probability

The probability $P_w = P_e + P_f$ is the probability for Algorithm 2 to yield a wrong decoding result, i.e., the probability for obtaining a decoding error or a decoding failure. For errors of weight $t \leq t_{\rm g}$, we know from Theorem 1 that we never fail or get an erroneous decoding result, and consequently $P_w(t) = 0$. For a received word $\boldsymbol{R}$ with $t > t_{\rm g}$ column errors, the probability $P_w(t)$ is simply the sum of the error probability $P_e(t)$ and the failure probability $P_f(t)$. Hence, $P_w(t)$ is overbounded by the sum $\overline{P}_e(t) + \overline{P}_f(t)$, or more precisely by

$$P_w(t) \leq \overline{P}_w(t) \triangleq \begin{cases} 0 & t \leq t_{\rm g} \\ \min\{\overline{P}_e(t) + \overline{P}_f(t), 1\} & t > t_{\rm g} \end{cases}. \quad (23)$$

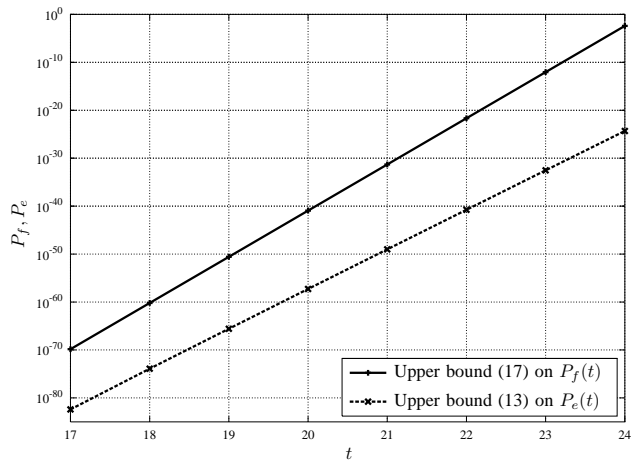

Fig. 3. Upper bounds on the failure and error probability for an IRS code composed of three codewords from the code $\mathcal{RS}(2^8; 255, 223, 33)$.

Now, we consider a codeword $\boldsymbol{A} \in \mathcal{A}$ with elements from $\mathbb{F}_q$, which is corrupted by an error matrix $\boldsymbol{E}$. We assume that a non-zero column in $\boldsymbol{E}$ occurs with probability $p$, and that the non-zero column vectors are distributed uniformly over $\mathbb{F}_q^l \setminus \{\boldsymbol{0}\}$. With this error model, the probability for $t$ column errors is

$$\binom{N}{t}p^t \cdot (1-p)^{N-t},$$

and we can overbound the overall word error probability by

$$P_w \leq \sum_{t=t_{\rm g}+1}^{N}\binom{N}{t} \cdot \overline{P}_w(t) \cdot p^t \cdot (1-p)^{N-t}. \quad (24)$$

## VII. IRS CODES IN CONCATENATED CODE DESIGNS

Reed–Solomon codes are commonly used as outer codes in concatenated code designs. In many applications, we even find schemes with interleaved Reed–Solomon codes of some kind. However, usually the Reed–Solomon codewords of such schemes are decoded independently, i.e., the special interleaved structure is typically not taken into account by the decoder.

To investigate the applicability of collaborative decoding strategies for IRS codes in concatenated code designs, we consider simple concatenated schemes with outer IRS codes and inner block codes.

## A. Concatenated Code Construction

To construct a concatenated code, we consider a codeword $\boldsymbol{A} \in \mathcal{A}$ of an IRS code over the field $\mathbb{F}_q$, where $q = p^m$. We group the elements of the codeword matrix

$$\boldsymbol{A} = \begin{pmatrix} \boldsymbol{a}^{(1)} \\ \vdots \\ \boldsymbol{a}^{(l)} \end{pmatrix} = (\boldsymbol{a}_0, \ldots, \boldsymbol{a}_{N-1}) \,,$$

into the column vectors

$$\boldsymbol{a}_j = \begin{pmatrix} a_j^{(1)} \\ \vdots \\ a_j^{(l)} \end{pmatrix}, \; j = 0, \ldots, N-1 \,.$$

Then, we select a linear code $\mathcal{B}$ of length $n$, dimension $k = lm$ and minimum distance $d$ over the field $\mathbb{F}_p$ and define a one-to-one mapping $g(\boldsymbol{a}) : \boldsymbol{a} \in \mathbb{F}_q^l \mapsto \boldsymbol{b} \in \mathcal{B}$. We apply this mapping to the column vectors $\boldsymbol{a}_j, \; j = 0, \ldots, N-1$, of the matrix $\boldsymbol{A}$. In this way, we obtain a matrix

$$\boldsymbol{C} = \left( g(\boldsymbol{a}_0), \ldots, g(\boldsymbol{a}_{N-1}) \right) = \left( \boldsymbol{b}_0^\mathsf{T}, \ldots, \boldsymbol{b}_{N-1}^\mathsf{T} \right) \,, \quad (25)$$

in which all columns are codewords of the inner code $\mathcal{B}$. We define the set $\{\boldsymbol{C}\}$ of all matrices obtainable in this way to be the concatenated code $\mathcal{C}$. For practical applications, usually it is preferable to have codes with a binary representation. Therefore, we consider binary inner codes, i.e., $p = 2$, and outer IRS codes over an extension field of $\mathbb{F}_2$, i.e., $q = 2^m$. We assume that the concatenated code $\mathcal{C}$ is used for transmission over an *Additive White Gaussian Noise* (AWGN) channel with *Binary Phase Shift Keying* (BPSK) modulation. In this case, a codeword

$$\boldsymbol{C} = \left( \boldsymbol{b}_0^\mathsf{T}, \ldots, \boldsymbol{b}_{N-1}^\mathsf{T} \right) = \left( c_{i,j} \right) ,$$

$c_{i,j} \in \mathbb{F}_2$, is mapped to the signal matrix

$$\boldsymbol{X} = \left( \boldsymbol{x}_0^\mathsf{T}, \ldots, \boldsymbol{x}_{N-1}^\mathsf{T} \right) = \left( x_{i,j} \right)$$

by $x_{i,j} = (-1)^{c_{i,j}}$, and transmitted over the AWGN channel. At the output, we observe the matrix

$$\boldsymbol{Y} = \left( \boldsymbol{y}_0^\mathsf{T}, \ldots, \boldsymbol{y}_{N-1}^\mathsf{T} \right) = \left( y_{i,j} \right) ,$$

with $y_{i,j} = x_{i,j} + n_{i,j}$. Since we use energy 1 to transmit a binary *code symbol*, the variables $n_{i,j}$ are statistically independent Gaussian random variables with zero mean and variance $\sigma^2 = N_0/2$, where $N_0$ is the single sided noise power spectral density. Moreover, for a concatenated code $\mathcal{C}$ of rate $R = \frac{Kk}{Nn}$, the effective energy for transmitting a single *information bit* is specified by the code rate $R$. Hence, we are able to characterize the AWGN channel by its signal-to-noise ratio, or more precisely by the ratio $\frac{E_b}{N_0} = \frac{1}{2R\sigma^2}$.

## B. Decoding of the Concatenated Code

To decode $\boldsymbol{Y}$ with respect to the inner code, we use a *Maximum Likelihood* (ML) decoder for the code $\mathcal{B}$ to find the ML estimates

$$\hat{\boldsymbol{b}}_j^\mathsf{T} = \arg\max_{\boldsymbol{b} \in \mathcal{B}} \left( P(\boldsymbol{b}^\mathsf{T} | \boldsymbol{y}_j^\mathsf{T}) \right) \quad (26)$$

for all columns of $\boldsymbol{Y}$. Then, we use the inverse mapping $g^{-1}(\boldsymbol{b}^\mathsf{T}) : \boldsymbol{b}^\mathsf{T} \in \mathcal{B} \mapsto \boldsymbol{a} \in \mathbb{F}_q^l$ to obtain the matrix

$$\boldsymbol{R} = \left( g^{-1}(\hat{\boldsymbol{b}}_0^\mathsf{T}), \ldots, g^{-1}(\hat{\boldsymbol{b}}_{N-1}^\mathsf{T}) \right) = \begin{pmatrix} \boldsymbol{r}^{(1)} \\ \vdots \\ \boldsymbol{r}^{(l)} \end{pmatrix} . \quad (27)$$

Each row $\boldsymbol{r}^{(\ell)}$, $\ell = 1, \ldots, l$, of $\boldsymbol{R}$ corresponds to an input word of the Reed–Solomon code $\mathcal{A}^{(\ell)}$. Hence, basically any row $\boldsymbol{r}^{(\ell)}$ can be decoded independently from all others with respect to the corresponding Reed–Solomon code $\mathcal{A}^{(\ell)}$. However, since an erroneous decision of the inner ML decoder may affect a complete column of the matrix $\boldsymbol{R}$, it occurs to be more expedient to apply the collaborative decoding strategy described in Section III to decode all rows of the IRS code simultaneously. In this case, the word error probability $p_w$ at the output of the inner decoder will be the column burst error probability at the input of the collaborative decoder for the outer IRS code. In other words, the probability $p_w$ is the probability that a column of $\boldsymbol{R}$ is erroneous.

Unfortunately, the exact analytical calculation of $p_w$ is generally not feasible. However, there are several known techniques to bound $p_w$ from both sides. One of the easiest and most widely known techniques to overbound the ML decoding error probability is the *Union Bound*.

## C. Error Probability after Inner ML Decoding

Unfortunately, the Union Bound is not very tight for channels with a bad to moderate signal-to-noise ratio, so that it is not suitable for analyzing our concatenated scheme. Therefore, we consider Poltyrev's *Tangential Sphere Bound* (TSB) [29]. Using this bound, $p_w$ can be overbounded by

$$p_w \leq \overline{p}_{\text{TSB}} \triangleq \sum_{w=1}^{\lceil \frac{t_0}{t_0+1} n - 1 \rceil} A_w \gamma(w, \sigma^2, t_0) + \zeta(\sigma^2, t_0) \,, \quad (28)$$

where

$$\gamma(w, \sigma^2, t_0) \triangleq \frac{2}{\pi} \exp\left( -\frac{n}{2\sigma^2} \right) \cdot$$

$$\cdot \int_0^\infty \int_{\sqrt{\frac{w}{n-w}}}^{\sqrt{t_0}} \frac{x}{1+y^2} \exp\left(-x^2\right) \cosh\left( 2x \sqrt{\frac{n}{2\sigma^2(1+y^2)}} \right) \cdot$$

$$\cdot \left( 1 - \overline{\Gamma}\left( \frac{n-2}{2}, \frac{t_0 - y^2}{1+y^2} x^2 \right) \right) dy \, dx \,,$$

$$\zeta(\sigma^2, t_0) \triangleq \frac{1}{\sqrt{\pi}} \int_{-\infty}^\infty \exp\left( -\left( x + \sqrt{\frac{n}{2\sigma^2}} \right)^2 \right) \cdot$$

$$\cdot \overline{\Gamma}\left( (n-1)/2, t_0 x^2 \right) dx \,,$$

and

$$\overline{\Gamma}(a, z) \triangleq \frac{1}{\Gamma(a)} \Gamma(a, z)$$

is the normalized *incomplete Gamma function*. The parameter $t_0$ is calculated from the cone half-angle

$$\theta_w = \arccos\left( \sqrt{\frac{w}{t_0(n-w)}} \right)$$

by solving the equation

$$\sum_{w=1}^{n-1} A_w \left(1 - u\left(w - \frac{t_0}{t_0+1}n\right)\right) \int_0^{\theta_w} \sin^{n-3}(\vartheta) d\vartheta = \frac{\Gamma((n-2)/2)}{\sqrt{\pi}\Gamma((n-1)/2)},$$

where

$$u(x) = \begin{cases} 0 & x < 0 \\ 1 & x \geq 0 \end{cases}$$

is the *Heaviside step function*.

To bound $p_w$ from below for channels with bad and moderate signal-to-noise ratios, Shannon's *Sphere Packing Bound* [30] is well suited for our purposes. The Sphere Packing Bound is calculated by

$$p_w \geq \underline{p}_{\text{SPB}} \triangleq \frac{1}{2^{\frac{n-1}{2}}\Gamma((n-1)/2)} \cdot \int_0^\infty x^{n-2} \exp\left(\frac{-x^2}{2}\right) \text{erfc}\left(\frac{\sqrt{\frac{n}{\sigma^2}} - \cot(\theta)x}{\sqrt{2}}\right) dx, \quad (29)$$

where the half angle $\theta$ is obtained by solving the equation

$$\int_0^\theta \sin^{n-2}(\vartheta) d\vartheta = \frac{\sqrt{\pi}\Gamma((n-1)/2)}{M\Gamma(n/2)},$$

with $M = |\mathcal{B}|$.

For good channels, i.e., high signal-to-noise ratios, the Sphere Packing bound is known to be rather weak. Therefore, we use Séguin's $L_2$–Bound [31] to bound $p_w$ from below for good channels. The $L_2$ bound is calculated by

$$p_w \geq \underline{p}_{L_2} \triangleq \frac{1}{2} \sum_{w=1}^n \frac{A_w \, \text{erfc}^2\left(\sqrt{\frac{w}{2\sigma^2}}\right)}{\Omega}. \quad (30)$$

At this, the denominator $\Omega$ is calculated by

$$\Omega \triangleq \text{erfc}\left(\sqrt{\frac{w}{2\sigma^2}}\right) + \\ + 2(A_w - 1)\psi\left(\alpha(w,w), \sqrt{\frac{w}{\sigma^2}}, \sqrt{\frac{w}{\sigma^2}}\right) + \\ + 2\sum_{\substack{v=1 \\ v \neq w}}^n A_v \psi\left(\alpha(w,v), \sqrt{\frac{w}{\sigma^2}}, \sqrt{\frac{w}{\sigma^2}}\right),$$

where

$$\alpha(i,j) \triangleq \min\left\{\sqrt{\frac{i}{j}}, \sqrt{\frac{j}{i}}, \frac{i+j-d}{2\sqrt{ij}}\right\},$$

and

$$\psi(\rho, x, y) \triangleq \\ \frac{1}{2\pi\sqrt{1-\rho^2}} \int_x^\infty \int_y^\infty \exp\left(-\frac{u^2 - 2\rho uv + v^2}{2(1-\rho^2)}\right) du dv$$

is the bivariate Gaussian function.

To obtain a lower bound on $p_w$, we take the maximum of the Sphere Packing bound (29), and the $L_2$ bound (30), i.e.

$$p_w \geq \underline{p}_w \triangleq \max\left\{\underline{p}_{\text{SPB}}, \underline{p}_{L_2}\right\}. \quad (31)$$

Since the TSB is quite tight for all channel conditions, it directly serves us as upper bound on $p_w$, i.e.,

$$p_w \leq \overline{p}_w \triangleq \overline{p}_{\text{TSB}}. \quad (32)$$

The bounds (31) and (32) turn out to be beneficial to derive bounds on the decoding performance after outer decoding. For decoding the outer code, we basically have two options. Either we apply Algorithm 2 for collaboratively decoding the outer IRS code, or we use a classical BMD decoder to decode the $l$ outer Reed–Solomon words independently row by row.

### D. Word Error Probability with Collaborative Outer Decoding

We first analyze the collaborative decoding strategy, which applies Algorithm 2 for decoding the outer code. To overbound the word error probability $P_w^c$ in this case, it would be helpful to apply the Theorems 5 and 6 derived in Section III. However, Theorem 6 requires that all column errors occur equiprobable in the IRS scheme. Unfortunately, this is not true after decoding the inner code, since due to the characteristics of ML decoding, low-weight error patterns occur more frequently than high-weight patterns. To be able to apply Theorem 6 anyway, we slightly modify the our concatenated coding scheme to randomize the error patterns after inner decoding. For this purpose, let $\mathcal{M}_l$ be the set of all non-singular $l \times l$ matrices with elements from the field $\mathbb{F}_q$. Now we modify the encoding rule given by Equation (25) into

$$\begin{aligned} \boldsymbol{C}' &= (g(\boldsymbol{M}_0 \cdot \boldsymbol{a}_0), \ldots, g(\boldsymbol{M}_{N-1} \cdot \boldsymbol{a}_{N-1})) \\ &= \left(\boldsymbol{\beta}_0^\mathsf{T}, \ldots, \boldsymbol{\beta}_{N-1}^\mathsf{T}\right), \end{aligned} \quad (33)$$

where the matrices $\boldsymbol{M}_i$ are statistically independent random matrices, uniformly distributed in $\mathcal{M}_l$. The reverse mapping after inner decoding described by Equation (27) is modified to

$$\boldsymbol{R} = \left(\boldsymbol{M}_0^{-1} \cdot g^{-1}(\hat{\boldsymbol{\beta}}_0^\mathsf{T}), \ldots, \boldsymbol{M}_{N-1}^{-1} \cdot g^{-1}(\hat{\boldsymbol{\beta}}_{N-1}^\mathsf{T})\right). \quad (34)$$

This randomization procedure does not influence the uncorrupted columns in $\boldsymbol{R}$, but only ensures the erroneous columns after inner decoding to be transformed into uniformly distributed error patterns. Since the number of erroneous columns is not changed by randomization, it should have no negative impact on the decoding performance. However, this randomization procedure allows us to apply Theorem 6 for estimating the failure probability after outer decoding. Later, we will observe by means of experimental results that from a practical point of view randomization is not necessary, since the decoding results will be virtually the same, regardless whether we use randomization or not.

**Theorem 7 (Upper Bound on $P_w^c$)** *Consider a concatenated code design with randomization as described above, and assume that the outer component codes are chosen such that (11) is satisfied. Furthermore, assume that the words of the* *inner code are decoded by a Maximum-Likelihood Decoder and the words of the outer interleaved Reed–Solomon code are decoded jointly by Algorithm 2. Then, the word error probability after decoding is overbounded by*

$$P_w^{\text{c}} \leq \sum_{t=t_{\text{g}}+1}^{N} \binom{N}{t} \cdot \overline{P}_w(t) \cdot \overline{p}_N\left(t, \overline{p}_w, \underline{p}_w\right) , \quad (35)$$

*where*

$$\overline{P}_w(t) \triangleq \min\left\{\overline{P}_f(t) + \overline{P}_e(t), 1\right\} ,$$

*and*

$$\overline{p}_N\left(t, \overline{p}_w, \underline{p}_w\right) = \begin{cases} \underline{p}_w^t \cdot (1-\underline{p}_w)^{(N-t)}, & i \leq \underline{p}_w N \\ \overline{p}_w^t \cdot (1-\underline{p}_w)^{(N-t)}, & \underline{p}_w N < t < \overline{p}_w N \\ \overline{p}_w^t \cdot (1-\overline{p}_w)^{(N-t)}, & t \geq \overline{p}_w N \end{cases}. \quad (36)$$

*Proof:* For proving Theorem 7, we consider the fact that the word error probability $p_w$ after inner decoding coincides with the column error probability $p$ at the input of the outer IRS decoder. Hence, according to (24) the word error probability $P_w^{\text{c}}$ after outer decoding is overbounded by

$$P_w^{\text{c}} \leq \sum_{t=t_{\text{g}}+1}^{N} \binom{N}{t} \cdot \overline{P}_w(t) \cdot p_w^t (1-p_w)^{N-t} , \quad (37)$$

where

$$\overline{P}_w(t) \triangleq \min\left\{\overline{P}_f(t) + \overline{P}_e(t), 1\right\} .$$

To further overbound (37), we define the function

$$f(p_w) \triangleq p_w^t \cdot (1-p_w)^{N-t},$$

and calculate the derivative

$$\tfrac{d}{dp_w} f(p_w) = t p_w^{t-1} (1-p_w)^{N-t} - p_w^t (N-i)(1-p_w)^{N-t-1}. \quad (38)$$

From this derivative we observe that $f(p_w)$ is monotonically non-decreasing as long as $p_w \leq \tfrac{t}{N}$, and monotonically non-increasing, if $p_w \geq \tfrac{t}{N}$. Consequently, to overbound $f(p_w)$, we replace $p_w$ by $\overline{p}_w$ if $t \leq \overline{p}_w N$, and by $\underline{p}_w$ if $t \geq \underline{p}_w N$. In the range $\underline{p}_w N < t < \overline{p}_w N$, we do not know, whether $f(p_w)$ is increasing or decreasing, so we overbound it by $f(p_w) \leq \overline{p}_w^t (1-\underline{p}_w)^{N-t}$. This results in the following upper bound:

$$\begin{aligned} f(p_w) &\leq \overline{p}_N\left(t, \overline{p}_w, \underline{p}_w\right) = \\ &= \begin{cases} \underline{p}_w^t \cdot (1-\underline{p}_w)^{(N-t)}, & t \leq \underline{p}_w N \\ \overline{p}_w^t \cdot (1-\underline{p}_w)^{(N-t)}, & \underline{p}_w N < t < \overline{p}_w N \\ \overline{p}_w^t \cdot (1-\overline{p}_w)^{(N-t)}, & t \geq \overline{p}_w N \end{cases}. \end{aligned}$$

Hence, replacing $f(p_w) = p_w^t \cdot (1-p_w)^{N-t}$ in (37) by $\overline{p}_N\left(t, \overline{p}_w, \underline{p}_w\right)$ yields (35), and consequently proves Theorem 7. ∎

### E. Word Error Probability with Independent Outer Decoding

Now, we consider the case that we decode each row of $\boldsymbol{R}$ independently by a classical BMD decoder. We overbound the word error probability $P_w^{\text{i}}$ for independent decoding, i.e., the probability that BMD decoding fails or is erroneous for at least one of the $l$ rows, by the following theorem:

**Theorem 8 (Upper Bound on $P_w^{\text{i}}$)** *Consider a concatenated code design as described above, and assume that the outer component codes are chosen such that (11) is satisfied. Furthermore, assume that the words of the inner code are decoded by a Maximum-Likelihood Decoder and that the words of the outer interleaved Reed–Solomon code are decoded independently by a BMD decoder. Then, the word error probability after decoding is overbounded by*

$$P_w^{\text{i}} \leq \sum_{t=t_{\text{g}}+1}^{N} \binom{N}{t} \cdot \overline{p}_N\left(t, \overline{p}_w, \underline{p}_w\right) , \quad (39)$$

*where $\overline{p}_N\left(t, \overline{p}_w, \underline{p}_w\right)$ is calculated by (36).*

*Proof:* If the number of erroneous columns in $\boldsymbol{R}$ is $t \leq t_{\text{g}}$, we always obtain a correct decoding result. To overbound $P_w^{\text{i}}$, we assume that for $t > t_{\text{g}}$ the independent decoding strategy based on standard BMD decoders never takes a correct decision. In other words, we simply replace $\overline{P}_w(t)$ in Theorem 7 by one, which directly yields the statement of Theorem 8. ∎

Theorem 8 allows us to estimate the worst case decoding performance for independent decoding of the outer Reed–Solomon codes. However, to be able to assess the potential of our collaborative decoding strategy, we also require a lower bound on $P_w^{\text{i}}$. For this purpose, we first consider the following lemma:

**Lemma 7** *Consider an IRS code according to Definition 3, and assume that we have $t$ column errors, uniformly distributed over $\mathbb{F}_q^l \setminus \{\boldsymbol{0}\}$. Further, assume that the Reed–Solomon codewords are decoded independently by a BMD decoder. Then, the probability that a corrupted IRS word is not decoded correctly, i.e., the probability that the BMD decoder fails to decode at least one Reed–Solomon word, is lower bounded by*

$$P_e^{\text{i}}(t) \geq \underline{P}_e^{\text{i}}(t) \triangleq 1 - \frac{\prod_{\ell=1}^{l} \sum_{i=0}^{t_{\text{g}}^{(\ell)}} \binom{t}{i} (q-1)^i}{(q^l - 1)^t} , \quad (40)$$

*where $t_{\text{g}}^{(\ell)} = \left\lfloor \tfrac{N-K^{(\ell)}}{2} \right\rfloor$.*

*Proof:* Consider the set $\mathcal{N}$ of all $l \times t$ matrices, such that the Hamming weight of row $\ell$, $\ell = 1, \ldots, l$ is smaller or equal to $t_{\text{g}}^{(\ell)}$. The cardinality of this set is calculated by

$$N_c^* \triangleq |\mathcal{N}| = \prod_{\ell=1}^{l} \sum_{i=0}^{t_{\text{g}}^{(\ell)}} \binom{t}{i} (q-1)^i .$$

Each matrix in $\mathcal{N}$ which has no all-zero columns, represents a correctable error pattern, i.e., the BMD decoder is able to decode

$$N_c \leq N_c^*$$

different error patterns with $t$ erroneous columns. The total number of matrices without any all-zero columns is $(q^l - 1)^t$. Hence, the probability for an error pattern with $t$ column errors to be decoded is

$$P_c^i(t) = \frac{N_c}{(q^l-1)^t} \leq \frac{N_c^*}{(q^l-1)^t}.$$

Consequently, the probability that an error pattern with $t$ column errors is not decoded correctly, is obtained by

$$P_e^i(t) = 1 - P_c^i(t) \geq 1 - \frac{N_c^*}{(q^l-1)^t} =$$
$$= 1 - \frac{\prod_{\ell=1}^{l} \sum_{i=0}^{t_g^{(\ell)}} \binom{t}{i}(q-1)^i}{(q^l-1)^t}.$$

∎

Lemma 7 enables us, to state and prove the following theorem:

**Theorem 9 (Lower Bound on $P_w^i$)** *Consider a concatenated code design as described above, and assume that the outer component codes are chosen such that (11) is satisfied. Furthermore, assume that the words of the inner code are decoded by a Maximum-Likelihood Decoder and that the words of the outer interleaved Reed–Solomon code are decoded independently by a BMD decoder. Then, the word error probability after decoding is bounded from below by*

$$P_w^i \geq \sum_{t=t_g+1}^{N} \binom{N}{t} \cdot \underline{P}_e^i(t) \cdot \underline{p}_N\left(t, \overline{p}_w, \underline{p}_w\right), \quad (41)$$

*where $\underline{p}_N\left(t, \overline{p}_w, \underline{p}_w\right)$ is calculated by*

$$\underline{p}_N\left(t, \overline{p}_w, \underline{p}_w\right) \triangleq \begin{cases} \overline{p}_w^t \cdot (1-\overline{p}_w)^{(N-t)}, & t \leq \underline{p}_w N \\ \underline{p}_w^t \cdot (1-\overline{p}_w)^{(N-t)}, & \underline{p}_w N < t < \overline{p}_w N \\ \underline{p}_w^t \cdot (1-\underline{p}_w)^{(N-t)}, & t \geq \overline{p}_w N \end{cases}.$$

*Proof:* Assume that an IRS codeword is corrupted by $t > t_g$ column errors. Then, according to Lemma 7 these errors cannot be decoded successfully by independent BMD decoding with a probability of at least $\underline{P}_e^i(t)$. Hence, to obtain a lower bound on $P_w^i$, we weight the probabilities

$$\binom{n}{t} \cdot p_w^t \cdot (1-p_w)^{n-t}$$

for having $t$ column errors by $\underline{P}_e^i(t)$, and sum them up from $t_g + 1$ to $N$. In this way, we obtain the bound

$$P_w^i \geq \sum_{t=t_g+1}^{N} \binom{N}{t} \cdot \underline{P}_e^i(t) \cdot p_w^t \cdot (1-p_w)^{N-t}. \quad (42)$$

As before in the proof of Theorem 7, we consider the function $f(p_w) = p_w^t \cdot (1-p_w)^{n-t}$ and its derivative (38), and find that $f(p_w)$ is monotonically non-decreasing for $p_w \leq \frac{t}{N}$ and monotonically non-increasing for $p_w \geq \frac{t}{N}$. This allows us to lower bound $f(p_w)$ by

$$f(p_w) \geq \underline{p}_N\left(t, \overline{p}_w, \underline{p}_w\right) =$$
$$= \begin{cases} \overline{p}_w^t \cdot (1-\overline{p}_w)^{(N-t)}, & t \leq \underline{p}_w N \\ \underline{p}_w^t \cdot (1-\overline{p}_w)^{(N-t)}, & \underline{p}_w n < t < \overline{p}_w N \\ \underline{p}_w^t \cdot (1-\underline{p}_w)^{(N-t)}, & t \geq \overline{p}_w N \end{cases}.$$

Replacing $f(p_w)$ in (42) by $\underline{p}_N\left(t, \overline{p}_w, \underline{p}_w\right)$ proves Theorem 9.
∎

## VIII. PERFORMANCE EVALUATION

The bounds described in the previous section allow for analytically estimating the probability $P_w$ of independent and collaborative decoding. To demonstrate the achievable gain by collaborative IRS decoding, to verify our analytic results, and to get a visual impression of the bounds described in the previous section, we complement our investigations by *Monte Carlo* simulations. Moreover, we compute the upper and lower bounds described in Section VII for two specific concatenated code designs to understand the asymptotic behavior of these bounds.

In the following, we consider the concatenated code $\mathcal{C}_1$, which is obtained from a homogeneous IRS code composed of $l = 2$ codewords of the code $\mathcal{A} = \mathcal{RS}\left(2^6; 63, 54, 10\right)$. The columns of this IRS code are encoded by the well known *Golay code* $\mathcal{B} = \mathcal{G}(23)$. This yields a code of length 1449, dimension 648, and rate 0.45 over the binary field $\mathbb{F}_2$.

The second code $\mathcal{C}_2$ is composed of a homogeneous IRS code created from $l = 3$ words of the Reed–Solomon code $\mathcal{A} = \mathcal{RS}\left(2^8; 255, 223, 33\right)$, and a binary inner $(30, 24, 4)$ code $\mathcal{B}$, obtained by doubly shorting the *Reed–Muller code* $\mathcal{RM}(3, 5)$. In this way, we obtain a binary code of length 7650, dimension 5352, and rate 0.7.

### A. Monte Carlo Simulations

For bad and moderate signal-to-noise ratios, the decoding performance can be analyzed by Monte Carlo simulations. We perform simulations for independent decoding and for collaborative decoding for the two codes $\mathcal{C}_1$ and $\mathcal{C}_2$. For collaborative decoding, we consider two variants: collaborative decoding with randomized errors like described in Section VII, and collaborative decoding without randomization. In this way, we obtain reliable results for block error probabilities larger or equal to $10^{-6}$. For smaller block error rates, reliable results can only be obtained by extensive computational efforts. To check the tightness of the upper bounds (39) and (35), we compare our simulation results with these bounds. The results of these investigations are shown in Fig. 4 for the code $\mathcal{C}_1$ and in Fig. 5 for the code $\mathcal{C}_2$.

For the code $\mathcal{C}_1$ we observe a gain of collaborative decoding of about $0.6\,\mathrm{dB}$, at a block error rate $P_w = 10^{-6}$. The other way round, for an $\frac{E_b}{N_0}$–ratio of about $4\,\mathrm{dB}$, collaborative decoding is about 100 times superior to independent decoding. The upper bounds (39) and (35) are approximately $0.2\,\mathrm{dB}$ worse than the actual decoding performance in the depicted

18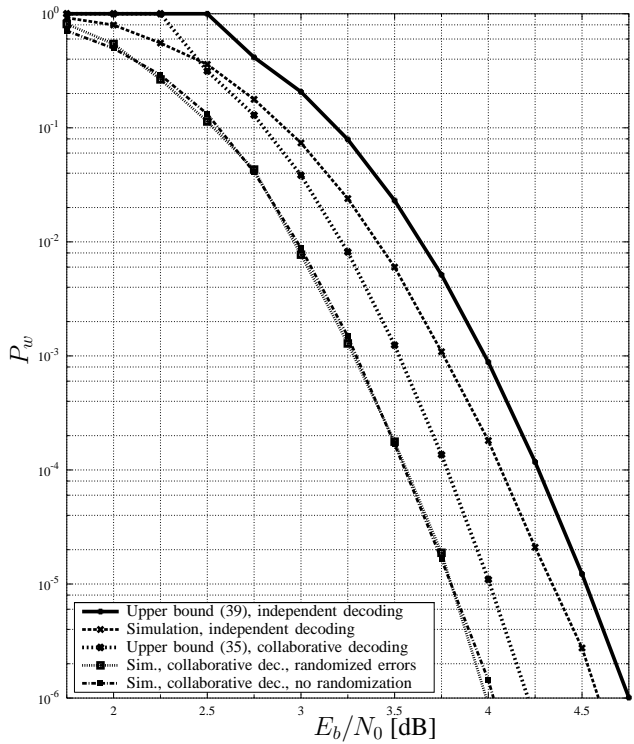

Fig. 4. Simulated decoding performance of the concatenated code $\mathcal{C}_1$ composed of two codewords from the outer code $\mathcal{RS}\left(2^6;63,54,10\right)$, and an binary inner Golay code $\mathcal{G}\left(23\right)$, AWGN channel, BPSK modulation.

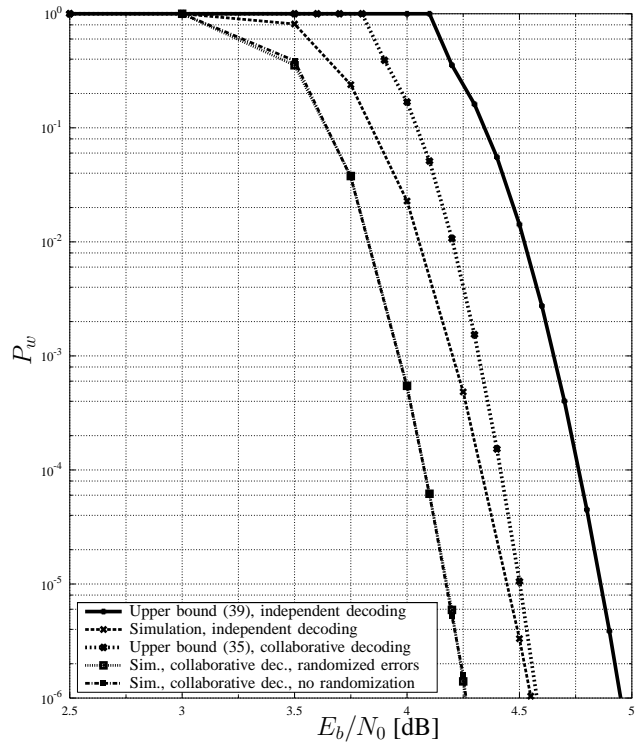

Fig. 5. Simulated decoding performance of the concatenated code $\mathcal{C}_2$ composed of three codewords from the outer code $\mathcal{RS}\left(2^8;255,223,33\right)$, and an binary linear inner $(30,24,4)$ code, AWGN channel, BPSK modulation.

range. However, the horizontal offset of the two bounds is also in the range of $0.5$ to $0.6$ dB, so that the bounds can be used for a rough estimation of the expected gain of collaborative decoding.

For the code $\mathcal{C}_2$, we observe a collaborative decoding gain of about $0.3$ dB at $P_w = 10^{-6}$, or in other words, the word error rate is about $1000$ times smaller with collaborative decoding compared to independent decoding. The upper bound (39) is about $0.4$ dB worse than the actual decoding performance for independent decoding, while the gap between the bound (35) and the collaborative decoding performance is about $0.3$ dB. Hence, an estimation of the collaborative decoding gain obtained by comparing (39) and (35) is a bit optimistic in this case.

We observe for both codes that we obtain virtually the same error probabilities $P_w$ for collaborative decoding with randomized errors, and for decoding without randomization. This means that from a practical point of view, randomization has no significant influence on the collaborative decoding performance. Hence, randomization is only necessary to fulfill the statistical requirements to derive a bound on the failure probability $P_f$.

## B. Asymptotic Considerations

Clearly, Monte Carlo simulations are only feasible, if the target block error probabilities are not too small. For applications which require very low block error rates like data storage systems or optical data transmission systems, we need analytical tools to estimate the obtainable decoding performance. The bounds derived in Section VII provide us with such tools, since they become tight for high signal-to-noise ratios. To see this, we calculate the bounds (41), (39), and (35) for the codes $\mathcal{C}_1$ and $\mathcal{C}_2$. The results for $\mathcal{C}_1$ are depicted by Fig. 6, and the results for $\mathcal{C}_2$ are depicted by Fig. 7.

We observe for both codes that the lower bound (41) for independent decoding is not very tight for channels with small signal-to-noise ratios, but converges to the upper bound (39) if the signal-to-noise ratio increases. For the code $\mathcal{C}_1$, the lower bound (41) and the upper bound (39) coincide beginning from an $\frac{E_b}{N_0}$–ratio of about $8$ dB, for $\mathcal{C}_2$ the two bounds coincide a little later at about $9$ dB. This means that asymptotically, i.e., for $\frac{E_b}{N_0} \to \infty$, we exactly know the performance of independent decoding.

Since we have the upper bound (35) on the collaborative error probability, and the lower bound (41) on the error probability of independent decoding, we are able to calculate the guaranteed collaborative decoding gain, i.e., the horizontal gap between (41) and (35). In Fig. 6 we observe that for $\mathcal{C}_1$ the collaborative decoding gain first increases, if the signal-to-noise ratio increases, and then vanishes again asymptotically, i.e., for $\frac{E_b}{N_0} \to \infty$. This behavior is explained by the fact that the asymptotic gain is specified by the guaranteed error correction radius, which in turn is specified by the minimum distance. In this sense, the asymptotic gain is rather a code property than a decoder property.

However, for all target block error rates which could be relevant for practical applications, we obtain a collaborative

PREPRINT -- PREPRINT -- PREPRINT -- PREPRINT

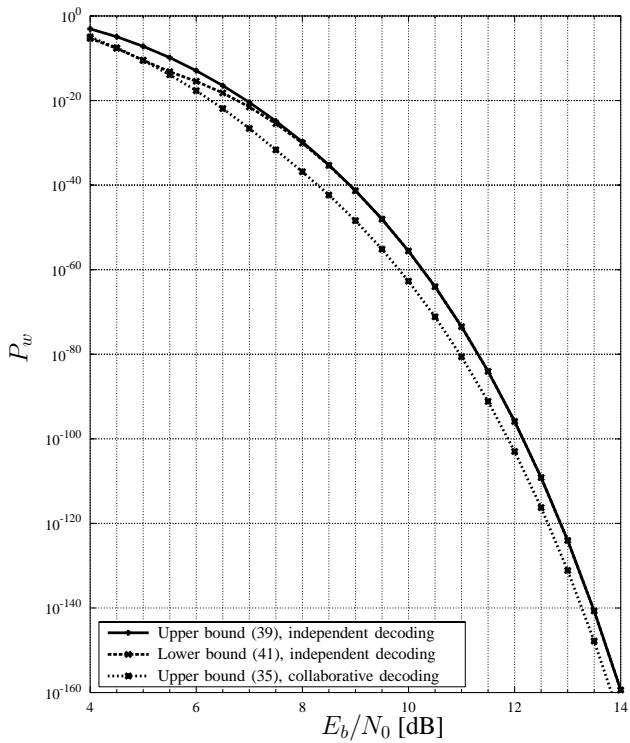

Fig. 6. Bounds on the decoding performance of the concatenated code $\mathcal{C}_1$ composed of two codewords from the outer code $\mathcal{RS}\left(2^6; 63, 54, 10\right)$, and an binary inner Golay code $\mathcal{G}\left(23\right)$, AWGN channel, BPSK modulation.

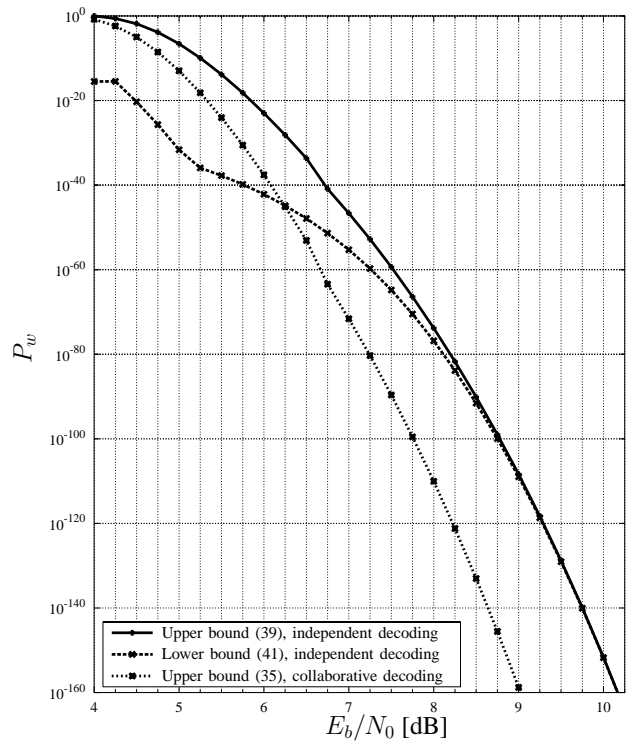

Fig. 7. Bounds on the decoding performance of the concatenated code $\mathcal{C}_2$ composed of three codewords from the outer code $\mathcal{RS}\left(2^8; 255, 223, 33\right)$, and an binary linear inner $(30, 24, 4)$ code, AWGN channel, BPSK modulation.

decoding gain by applying Algorithm 2. In Fig. 7 we see that the collaborative decoding gain, which can be guaranteed by our bounds, grows to more than $1\,\text{dB}$, when the signal-to-noise ratio improves to 9 or $10\,\text{dB}$. Since the computational complexity of Algorithm 2 is of the same order as the computational complexity of the Berlekamp–Massey algorithm, this gain can be obtained virtually for free.

## IX. CONCLUSIONS

In this paper, we presented a decoding algorithm for IRS codes, which is based on varying length multi-sequence shift-register synthesis. This algorithm allows for efficiently decoding both homogeneous and heterogeneous IRS codes. We explained that homogeneous IRS constructions yield MDS codes with an optimum minimum distance, while heterogeneous IRS code designs provide a high degree of freedom for constructing generalized concatenated codes and other special applications like the decoding of low rate Reed–Solomon codes as explained in [15].

The decoding of heterogeneous IRS codes is closely related to joint error and erasure correction. Hence, it is also possible to adopt our algorithm for performing joint error and erasure decoding.

The decoder described by Algorithm 2 is able to correct all error patterns within a sphere of radius $t_\text{g}$, where $t_\text{g}$ is half the minimum distance of the IRS code. Furthermore, it is able to correct errors beyond half the minimum distance, as long as the number of errors is below the maximum error correcting radius

$$t_\text{max} = \frac{l}{l+1}(N - \bar{K})\,.$$

If the decoding radius is increased beyond half the minimum distance, it is in principle not possible to decode all error patterns uniquely, since the received vector may lie in a region, where several correction spheres are overlapping. In this case, three basic strategies are conceivable: the decoder selects one solution from the list, if several solutions exist; the decoder yields all possible solutions and leaves it to the succeeding data processing unit to cope with this list; or the decoder does not take a decision at all and yields a decoding failure. Our algorithm applies the third strategy. Whenever it is not able to find a unique solution for a received word with $t$ errors, it declares a decoding failure. Clearly, this strategy makes only sense, if the failure probability is small enough, and hence most of the error patterns can be corrected in the range $t_\text{g} \leq t \leq t_\text{max}$. Therefore, we overbounded the failure probability $P_f(t)$ and showed that it is in the order of $\frac{1}{q}$, for $t = t_\text{max}$, and decreases exponentially with decreasing $t$. Besides of the failure probability $P_f$, we also overbounded the error probability $P_e$ to be able to analyze the overall performance of collaborative decoding, or more precisely the probability $P_w = P_e + P_f$ for a wrong decoding result.

Moreover, we considered concatenated code designs with outer IRS codes and inner block codes, since inner decoding causes column burst errors, which are necessary for effective collaborative decoding. We applied the bound on $P_e(t)$ and $P_f(t)$ to analyze the decoding performance of such concate-nated schemes. This analysis has been performed by deriving bounds on the word error probability after independent and collaborative outer decoding, and by using them to investigate, which decoding gains can be achieved by collaborative decoding in comparison to an independent decoding strategy. We complemented these considerations by Monte Carlo simulations for two specific code designs. We observed that for all operating points which could be interesting for practical applications, we are able to achieve a collaborative decoding gain without increasing the decoding complexity in comparison to independent decoding. The probabilities $P_w$ achievable by our collaborative decoding strategy are 100–1000 times smaller than the probabilities achievable by independent decoding.

The concatenated code designs considered here are rather simple, to be able to analyze them theoretically. However, since concatenated codes with interleaved Reed–Solomon codes are used in many applications and can be found in several standards, the basic principles discussed in this paper may also be applicable for such practical systems.

ACKNOWLEDGMENT

We would like to thank Christian Senger for many valuable comments on this manuscript.REFERENCES

[1] V. Y. Krachkovsky and Y. X. Lee, "Decoding for interleaved Reed-Solomon schemes," *Trans. Magn.*, vol. 33, pp. 2740–2743, September 1997.
[2] V. Y. Krachkovsky, "Reed–Solomon codes for correcting phased error bursts," *IEEE Trans. Inform. Theory*, vol. IT-49, pp. 2975–2984, November 2003.
[3] V. Y. Krachkovsky, Y. X. Lee, and H. K. Garg, "Decoding of parallel RS codes with applications to product and concatenated codes," in *Proc. IEEE Intern. Symposium on Inf. Theory*, (Boston, USA), p. 55, 1998.
[4] D. Bleichenbacher, A. Kiayias, and M. Yung, "Decoding of interleaved Reed Solomon codes over noisy data," in *Springer Lecture Notes in Computer Science*, vol. 2719, pp. 97–108, January 2003.
[5] A. Brown, L. Minder, and A. Shokrollahi, "Probabilistic decoding of interleaved RS-codes on the $q$-ary symmetric channel," in *Proc. IEEE Intern. Symposium on Inf. Theory*, (Chicago, IL, USA), p. 327, 2004.
[6] A. Brown, L. Minder, and A. Shokrollahi, "Improved decoding of interleaved AG codes," in *Cryptography and Coding*, vol. 3796 of *Lecture Notes in Computer Science*, pp. 37–46, Berlin: Springer Verlag, December 2005.
[7] J. Justesen, C. Thommesen, and T. Høholdt, "Decoding of concatenated codes with interleaved outer codes," in *Proc. IEEE Intern. Symposium on Inf. Theory*, (Chicago, IL, USA), p. 329, 2004.
[8] F. Parvaresh and A. Vardy, "Multivariate interpolation decoding beyond the Gurswami–Sudan radius," in *Proc. 42nd annual Allerton Conf. Commun., Control and Comput.*, (Urbana, IL, USA), 2004.
[9] G. Schmidt, V. R. Sidorenko, and M. Bossert, "Error and erasure correction of interleaved Reed–Solomon codes," in *Proc. Int. Workshop on Coding and Cryptography*, (Bergen, Norway), pp. 20–29, March 2005.
[10] G. Schmidt, V. R. Sidorenko, and M. Bossert, "Interleaved Reed–Solomon codes in concatenated code designs," in *Proc. IEEE ITSOC Inform. Theory Workshop*, (Rotorua, New Zealand), pp. 187–191, August 2005.
[11] G. Schmidt, V. R. Sidorenko, and M. Bossert, "Error and erasure correction of interleaved Reed–Solomon codes," in *Coding and Cryptography*, vol. 3969 of *Lecture Notes in Computer Science*, pp. 22–35, Berlin: Springer Verlag, 2006.
[12] G. Schmidt, V. R. Sidorenko, and M. Bossert, "Heterogeneous interleaved Reed–Solomon code designs," in *Proc. 10th Int. Workshop on Algebraic and Combinatorial Coding Theory (ACCT-10)*, (Zvenigorod, Russia), pp. 230–233, September 2006.
[13] E. L. Blokh and V. V. Zyablov, "Coding of generalized concatenated codes," *Problems of Information Transmission*, vol. 10, pp. 218–222, July–September 1974. Translated from Russian, original in Problemy Peredachi Informatsii, pp. 45–50.
[14] V. A. Zinoviev, "Generalized cascade codes," *Problems of Information Transmission*, vol. 12, pp. 2–9, January–March 1976. Translated from Russian, original in Problemy Peredachi Informatsii, pp. 5–15.
[15] G. Schmidt, V. R. Sidorenko, and M. Bossert, "Decoding Reed–Solomon codes beyond half the minimum distance using shift-register synthesis," in *Proc. IEEE Intern. Symposium on Inf. Theory*, (Seattle, WA, USA), pp. 459–463, July 2006.
[16] E. R. Berlekamp, *Algebraic Coding Theory*. New York: McGraw–Hill, 1968.
[17] J. L. Massey, "Shift-register synthesis and BCH decoding," *IEEE Trans. Inform. Theory*, vol. IT-15, pp. 122–127, January 1969.
[18] G.-L. Feng and K. K. Tzeng, "An iterative algorithm of shift-register synthesis for multiple sequences," *Scientia Sinica (Science in China), Series A*, vol. XXVIII, pp. 1222–1232, November 1985.
[19] G.-L. Feng and K. K. Tzeng, "A generalization of the Berlekamp-Massey algorithm for multisequence shift-register synthesis with applications to decoding cyclic codes," *IEEE Trans. Inform. Theory*, vol. IT-37, pp. 1274–1287, September 1991.
[20] G. Schmidt and V. R. Sidorenko, "Linear shift-register synthesis for multiple sequences of varying length." Preprint, available online at ArXiv, arXiv:cs.IT/0605044, 2006.
[21] G. Schmidt and V. R. Sidorenko, "Multi-sequence linear shift-register synthesis: The varying length case," in *Proc. IEEE Intern. Symposium on Inf. Theory*, (Seattle, WA, USA), pp. 1738–1742, July 2006.
[22] W. W. Peterson, "Encoding and error–correction procedures for the Bose–Chaudhuri codes," *IEEE Trans. Inform. Theory*, vol. IT-6, pp. 459–470, September 1960.
[23] W. W. Peterson, *Error–correcting Codes*. Cambridge: M.I.T. Press, 1961.
[24] R. E. Blahut, *Theory and Practice of Error Control Codes*. Reading, Massachusetts: Addison-Wesley Publishing Company, 1983. ISBN 0-201-10102-5.
[25] G. D. Forney, "On decoding BCH codes," *IEEE Trans. Inform. Theory*, vol. IT-11, pp. 549–557, October 1965.
[26] R. E. Blahut, "Transform techniques for error control codes," *IBM J. Research and Development*, vol. 23, pp. 299–315, May 1979.
[27] T. Schaub, *A Linear Complexity Approach to Cyclic Codes*. PhD thesis, Swiss Federal Institute of Technology, Zürich, 1988.
[28] F. J. MacWilliams and N. J. Sloane, *The Theory of Error Correcting Codes*. Amsterdam: North-Holland, 1992. ISBN 0-444-85193-3.
[29] G. Poltyrev, "Bounds on the decoding error probability of binary linear codes via their spectra," *IEEE Trans. Inform. Theory*, vol. IT-40, pp. 1284–1292, July 1994.
[30] C. E. Shannon, "Probability of error for optimal codes in a Gaussian channel," *Bell Syst. Tech. J.*, vol. 38, pp. 611–656, May 1959.
[31] G. E. Séguin, "A lower bound on the error probability for signals in white Gaussian noise," *IEEE Trans. Inform. Theory*, vol. IT-44, pp. 3168–3174, November 1998.
PREPRINT -- PREPRINT -- PREPRINT -- PREPRINT